\documentclass[%
 reprint,
 superscriptaddress,
 amsmath,amssymb,
 aps,
]{revtex4-2}

\usepackage{graphicx}
\usepackage{dcolumn}
 \usepackage{bm}
\usepackage{hyperref}
\hypersetup{
	colorlinks=true,
	linkcolor=blue,
	filecolor=magneta,      
	urlcolor=blue,
}
\usepackage{amsmath}
\usepackage{microtype}
\usepackage{tabularx}
\usepackage{slashbox}
\usepackage[caption=false]{subfig}
 
\newcommand{\comment}[1]{}
\usepackage[usenames,dvipsnames]{xcolor}

\newcommand{\reals}{ \mathbb{R} }
\newcommand{\Dataset}{ \mathcal{D} }
\newcommand{\orcid}[1]{\href{https://orcid.org/#1}{\textcolor[HTML]{A6CE39}{\aiOrcid}}}

\begin{document}

\preprint{APS/123-QED}

\title{Parameters Estimation for the Cosmic Microwave Background \\ with Bayesian Neural Networks}

\author{H\'ector J. Hort\'ua\href{https://orcid.org/0000-0002-3396-2404}{\includegraphics[scale=0.5]{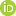}}}
\email{hortua.orjuela@rist.ro}
\author{Riccardo Volpi\href{https://orcid.org/0000-0003-4485-9573}{\includegraphics[scale=0.5]{ORCID-iD_icon-16x16.png}}}%
\email{volpi@rist.ro}
\affiliation{%
Machine Learning and Optimization Group,\\
Romanian Institute of Science and Technology (RIST),\\
Cluj-Napoca, Romania
}

\author{Dimitri Marinelli}
\email{dm@financial-networks.eu}
\affiliation{
 FinNet, Frankfurt am Main, Germany.  \\
}
\author{Luigi Malag\`o}%
\email{malago@rist.ro}
\affiliation{%
Machine Learning and Optimization Group,\\
Romanian Institute of Science and Technology (RIST),\\
Cluj-Napoca, Romania
}

%


\date{\today}

\begin{abstract}
In this paper, we present the first study that compares different models of Bayesian Neural Networks (BNNs) to predict the posterior distribution of the cosmological parameters directly from the Cosmic Microwave Background temperature and polarization maps.
We focus our analysis on four different methods to sample the weights of the network during training: Dropout, DropConnect, Reparameterization Trick (RT), and Flipout.
We find out that  Flipout outperforms all other methods regardless of the architecture used, and provides tighter constraints for the cosmological parameters.
Moreover we compare with  MCMC posterior analysis obtaining comparable error correlation among parameters, with BNNs being
orders of magnitude faster in inference, although less accurate.
Thanks to the speed of the inference process with BNNs, the posterior distribution, outcome of the neural network, can be used as the initial proposal for the Markov Chain. We show that this combined approach increases the acceptance rate in the Metropolis-Hasting algorithm and accelerates the convergence of the MCMC, while reaching the same final accuracy. In the second part of the paper, we present a guide to the training and calibration of a successful multi-channel BNN for the CMB temperature and polarization map. We show how tuning the regularization parameter for the standard deviation of the approximate posterior on the weights in Flipout and RT we can produce unbiased and reliable uncertainty estimates, i.e., the regularizer acts like a hyperparameter analogous to the dropout rate in Dropout. The best performances are nevertheless achieved with a more convenient method, in which the network parameters are let free during training to achieve the best uncalibrated performances, and then the confidence intervals are calibrated in a subsequent phase. Additionally, we describe existing strategies for calibrating the networks and propose new ones.
Finally, we show how polarization, when combined with the temperature in a unique multi-channel tensor fed to a single BNN, helps to break degeneracies among parameters and provides stringent constraints.
The results reported in the paper can be extended to other cosmological datasets in order to capture features that can be extracted directly from the raw data, such as non-Gaussianity or foreground emissions.\\
\end{abstract}

\maketitle

\section{Introduction}
The Cosmic Microwave Background (CMB) is by far one of the most powerful datasets available in cosmolo\-gy  for  understanding the Universe \cite{doi:10.1142/6730,naselsky2006physics}.
Measurements within the last decade have yielded
strong support for the standard cosmological  spatially-flat $\Lambda$CDM model and provided precise  estimates  for  its  cosmological parameters~\cite{ refId0,Giovannini:2007xh,dodelson2003modern}.
This base model is  described by six  parameters divided into two groups: the {\it primordial} given by ($n_s$, $A_s$) that describe the initial state of the perturbations produced by quantum fluctuation during inflation,  and the  {\it late-time} group formed by $(\omega_c,\omega_b,\tau,\theta_{MC})$  which trace the linear evolution of the perturbations after re-entering the Hubble radius \cite{Aghanim:2018eyx,2018arXiv180706211P}.
In addition to the standard cosmological model, other parameters might provide a wealth of new information on cosmolo\-gy, e.g., the total mass of neutrinos, the effective extra relativistic degrees of freedom, the tensor-to-scalar ratio, non-Gaussianity parameters, among the others~\cite{PhysRevD.97.123544,PhysRevD.97.123507,PITROU20181,Hort_a_2017,Shiraishi_2013}.
Such parameters have been of great interest for cosmologists because they could produce significant departures from
the standard model and represent new physics in the early Universe.
Combining the next-generation of CMB  experiments along with large scale structure (LSS) probes will be the next step toward a precision  cosmolo\-gy  that
will allow us to constrain these  fundamental physics parameters and find out extensions to the $\Lambda$CDM model
\footnote{\url{http://www.cfhtlens.org/}\\
\url{http://kids.strw.leidenuniv.nl/index.php}\\
\url{http://www.darkenergysurvey.org}\\
\url{http://hsc.mtk.nao.ac.jp/ssp/}\\
\url{http://www.lsst.org}\\
\url{http://sci.esa.int/euclid/}\\
\url{http://wfirst.gsfc.nasa.gov}\\
\url{https://cmb-s4.org/}\\
\url{http://www.litebird-europe.eu/}\\
\url{http://www.core-mission.org/}\\
\url{https://simonsobservatory.org/}}.
However, the combination of these probes will also require more advanced statistical methods to analyze the dataset and an enormous computational effort. In fact, the estimation of the cosmological parameters demands the calculation of theoretical power spectra which are obtained through Einstein-Boltzmann Solvers (EBS)  like CLASS \footnote{\url{http://class-code.net}}\cite{Blas_2011}. Usually, these codes require few seconds  for computing the observables, depending on the complexity of the cosmological model.
Afterward, a comparison between the  predictions  at various points in the parameter space with the available observations is done, and based on the likelihood a best-fit of the parameters is obtained. Packages like cobaya~\cite{torrado2020cobaya} or montepython \footnote{\url{http://baudren.github.io/montepython.html}} use MCMC algorithms to sample from the posterior distribution and  fulfil  this task \cite{doi:10.1080/00107510802066753}.
However, this process is  computationally expensive for theoretical models that include large amount of parameters or contain ``slow parameters'' (most of them are late-time and delay the calculation of the power spectrum), since the EBS is  executed at each step in the parameter space. 

In recent years, deep neural networks have been used successfully in the field of cosmology as a way to confront the upcoming computational challenges.
Originally inspired by neurobiology, deep neural network models have become a powerful tool in machine learning due to their capacity of approximating functions   and   dynamics   by   learning   from   examples \cite{2019arXiv190204704K,Lin2017}.
Using deep learning methods as emulators for computing the cosmological observables has become a very popular application in cosmology.
Different authors have proposed to implement deep neural networks emulators for the EBS  \footnote{\url{https://github.com/marius311/pypico}}\cite{Fendt:2006uh}, either totally or even partially, i.e., only in places where traditional estimations are more time-consuming \cite{2019arXiv190705764A,2019arXiv190705881M}.
Deep neural networks have also been used for extracting the observables  directly from the raw data without requiring the power spectrum or other compressed information.  
Based on this strategy, deep learning has been employed in classification tasks for
detecting strongly lensed systems~\cite{Lanusse:2017vha,2017MNRAS.472.1129P},  discriminating cosmological models~\cite{Schmelzle:2017vwd,Perraudin:2018rbt}, or  detecting cosmic strings in the CMB maps \cite{Ciuca2017ACN}. Additionally, for regression tasks  deep learning provides a way to make inference  either in
 gravitational lensing systems \cite{Hezaveh:2017sht,PerreaultLevasseur:2017ltk},
 weak lensing or LSS data \cite{Gupta:2018eev,2019arXiv190203663R,Fluri:2018hoy},  reionization and 21cm observations~\cite{2019arXiv190404106D,2019MNRAS.484..282G},
 and CMB data~\cite{he2018analysis,CALDEIRA2019100307,2019arXiv190204083K,doi:10.1002/asna.201512351}, also in  generative models as a powerful alternative to cosmological numerical simulations~\cite{He13825}.
However, the use of deep neural networks may rise some problems. Indeed neural networks are prone to over-fitting, so analyzing the results only based on   point estimates might produce unreliable  predictions with spuriously high confidence~\cite{KWON2020106816,DBLP:journals/corr/PereyraTCKH17}. Therefore, the following question naturally arises: {\it how can we be sure that our model is certain about its outcomes?} This fundamental concern has been an object of study in the machine learning community and one of the most attractive approaches to address this issue relies on the use of Bayesian Neural Networks (BNNs)~\cite{Gal2015Dropout}. BNNs represent the probabilistic version of the traditional neural networks capturing the posterior probability of the outcomes and estimating their  predictive uncertainties.
One of the most popular techniques used to obtain the uncertainties in the Bayesian framework is called Dropout. Initially, Dropout was proposed in~\cite{JMLR:v15:srivastava14a} as a regularisation scheme, subsequently the authors in~\cite{Gal2015Dropout} developed a theoretical framework in which Dropout in neural networks can be interpreted as approximate  variational  inference for  deep Gaussian processes.
Applications of BNNs using Dropout in cosmology are shown in~\cite{PerreaultLevasseur:2017ltk,2018arXiv180800011M,he2018analysis}. Recent studies (see~\cite{wen2018flipout} and references therein) have found that other techniques can remarkably improve the performances of BNNs and  reduce the variance of the estimates.
Furthermore, in~\cite{Gal2015BayesianCN} the authors have claimed that Dropout  fails in some  architectures while others have discussed the reliability of this method~\cite{2018arXiv180603335O}.
Moreover, in general neural networks predictions suffer from a poor calibration over their  uncertainty estimations and tend to be overconfident in their predictions,  i.e., predicted posterior distributions do not reflect  actual correctness probabilities~\cite{Guo:2017:CMN:3305381.3305518}.  Different strategies  and metrics have been proposed  to calibrate these networks and  to be able to  evaluate the accuracy of the obtained uncertainties,  some of these methods will be analyzed in this paper. Based on the aforementioned discussion, our goal is twofold. Firstly, we want to show an appealing application of deep learning in cosmology by estimating the posterior distribution of the cosmological parameters  directly from simulated CMB maps. Secondly, we want to describe different techniques employed to generate reliable uncertainty estimates of the predicted parameters and discuss their performance for the CMB dataset. The paper is organized as follows. First, in Sec.~\ref{sec:MCMC}, we present a motivating example for the use of Bayesian Neural Networks in Cosmology. We present a fair comparison between BNNs and MCMC posterior analysis, and we show how these approaches can also be combined to obtain the best of both worlds. In the following part of the paper we present a guide to the training and calibration of a successful BNN for the CMB temperature and polarization map. In Sec.~\ref{sectII} we describe the two sources of uncertainty in Bayesian neural networks: aleatoric and epistemic, and their importance for quantifying confidence intervals. In Sec.~\ref{sectIII} we briefly summarize the framework of variational inference, and how to produce estimates of the uncertainties and correlations of the physical parameters.
Sec.~\ref{sectIV} contains a description of different methods used in the literature to approximate the posterior over the weights of the networks. Some of them  have been frequently used because of their simplicity of implementation, while other more recent techniques lead to better interpretation under the Bayesian framework and to improved performance.
We then describe the generation of the synthetic maps  used to train the inference models, including the  network architectures, in  Sec.~\ref{sectV}.
In Sec.~\ref{sec:calibration} we discuss the calibration methods  used to assess the reliability of uncertainty estimates, and in  Sec.~\ref{sectVII} we show our main results related to use of BNNs in a cosmological context, as well as the credible cosmological parameter contours for our model. Furthermore, we show how the results are improved when we include polarization maps as additional channels in the images and we describe preliminary  methods for network calibrations.  Finally, we
present our conclusions and final remarks in Sec.~\ref{sectVIII}.
\section{\label{sec:MCMC} Bayesian inference problem: MCMC and Variational Inference}
\begin{figure}[h!]
\begin{center}
\includegraphics[width=0.48\textwidth]{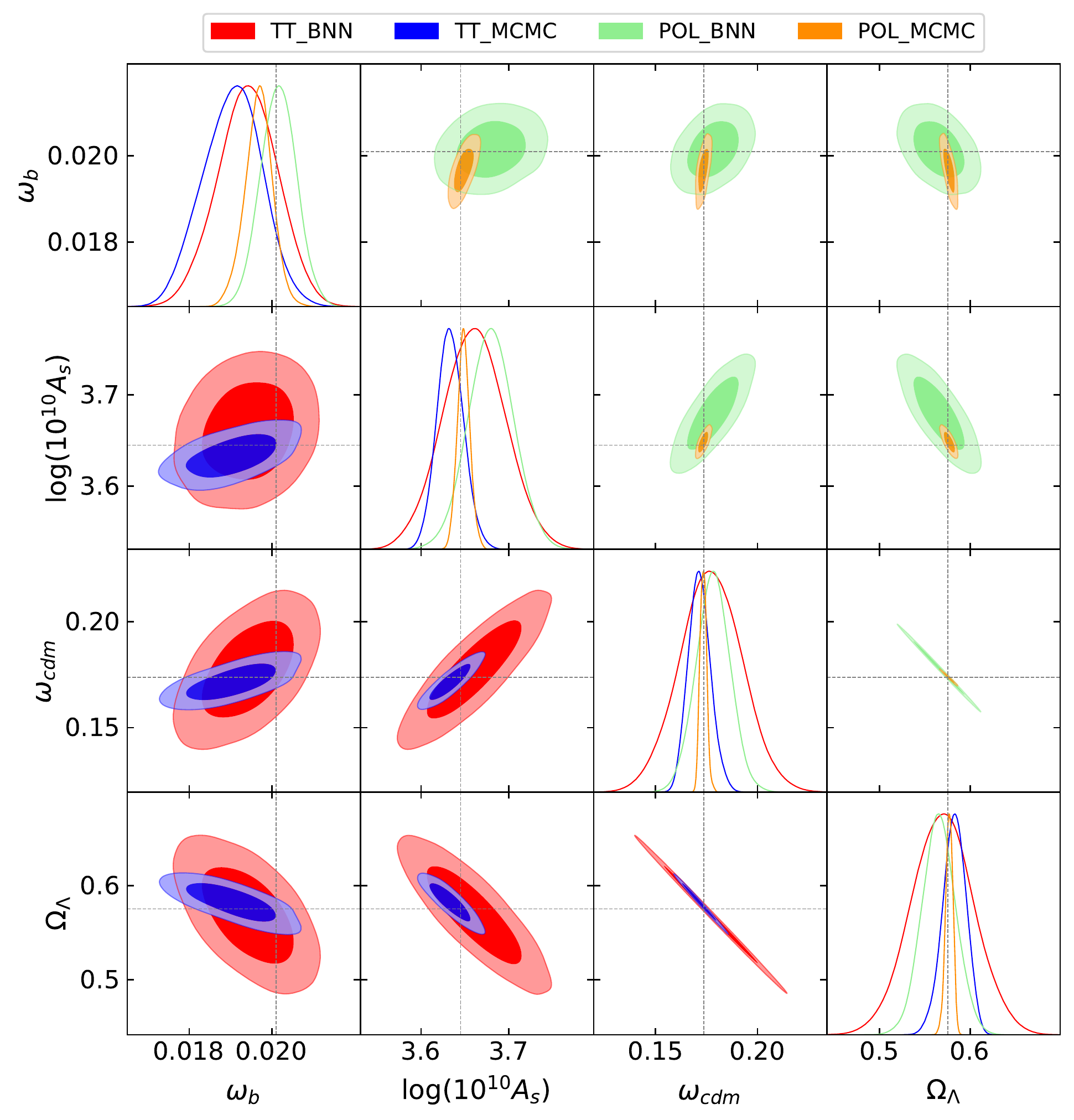}
\end{center}
\caption{\it  Marginalized  parameter  constraints  obtained from temperature maps (TT) and combined temperature  with polarization (POL) using MCMC and the best Bayesian Neural Network  (BNN) model reported in this paper. The latter model is calibrated using the method proposed in  Sec.~\ref{secpolcal}. The black line stand for the real value: $\omega_b=0.0201$, $\log(10^{10}A_s)=3.6450$ and  $\omega_{cdm}=0.1736$ taken from the test dataset. } \label{fig:mcmc}
\end{figure}

Bayesian  inference  offers a way for learning from data through the posterior distribution $p(\theta|d)\sim p(d|\theta)p(\theta)$;  being $\theta$  a set  of  unknown  parameters  of interest,  $d$ the data associated with a measurement, $p(\theta)$ the  prior distribution  that quantifies what we know about $\theta$ before observing any data, and  $p(d|\theta)$  is the likelihood  function.
Computing the true posterior is generally intractable, and approximation methods must be implemented in order to perform Bayesian inference in practice. Two main techniques  for  this  purpose  are Variational Inference and Markov Chain Monte Carlo (MCMC)~\cite{NIPS2011_4329,doi:10.1063/1.1699114,regier2018approximate,jain2018variational}. The former  method although is computationally faster, it requires  the  approximation  of  the  true  posterior (see detailed description  in Sec.~\ref{sectIII}).  The latter  has become one of the most popular  methods for cosmological parameter estimation due  to  its  advantage of being non-parametric and asymptotically exact.
Classical  MCMC methods  draw samples sequentially according to a probabilistic algorithm that allows to scale linearly with the dimension of the parameter space~\cite{verde2007practical}. However if the complexity of the model increases either by the presence of "slow" parameters, nuisance parameters related to instrument beam response, foregrounds or parameter correlations,  the sampling will exhibit a high numerical  cost~\cite{PhysRevD.87.103529}. Additionally,  it is generally difficult to determine a convenient initial state for the system and an accurate criterion to determine the convergence of the Markov Chain.
These practical issues compel MCMC practitioners to resort on MCMC convergence diagnostic tools and having to wait a long time to obtain good solutions~\cite{10.1111/j.1365-2966.2004.08464.x}.
In order to make a fair  comparison between both inference procedures, we  will use synthetic images (small patches of CMB maps) drawn from our model described in Sec.~\ref{sectV}. While MCMC takes in input the power spectrum extracted from the CMB maps (as it is standard practice in  cosmology), VI estimates the cosmological parameters directly from the raw maps themselves.  
This allows the Network to adaptively extract complicated correlations when performing inference without assuming a priori  summary statistics  such as  power spectrum or  higher order spectra (such as bispectrum, trispectrum or others). For MCMC sampling we use the package cobaya, with the likelihood given by~\cite{verde2007practical}
\begin{eqnarray}
  -\mathcal{L} &\sim& \sum_l(2l+1)\bigg[\ln\bigg(\frac{C_l^{BB}}{\hat{C}_l^{BB}}\bigg(\frac{C_l^{TT}C_l^{EE}-(C_l^{TE})^2}{\hat{C}_l^{TT}\hat{C}_l^{EE}-(\hat{C}_l^{TE})^2}\bigg)\bigg)\nonumber\\
  &+& \frac{\hat{C}_l^{BB}}{C_l^{BB}}+\frac{\hat{C}_l^{TT}C_l^{EE}+C_l^{TT}\hat{C}_l^{EE} -2\hat{C}_l^{TE}C_l^{TE}}{C_l^{TT}C_l^{EE}-(C_l^{TE})^2}\bigg]\,,\,
\end{eqnarray}
where $\hat{C}_l$ is the power spectrum of the CMB map, in case of the full-sky is obtained from healpy~\cite{Zonca2019}, while in case of a  patch is obtained by azimuthal averaging in Fourier space with Lens-Tools~\cite{2016A&C....17...73P}, and
$C_l$ is the theoretical model. Cobaya accepts the cosmological parameters as input, compute $C_l$ via CLASS and when the Markov chains have enough points to provide reasonable samples from the  posterior distributions, the simulation stops and it  return  the chains. The results are displayed in Fig.~\ref{fig:mcmc} where we compared the MCMC results with the best calibrated BNN model reported in  Sec.~\ref{secpolcal} (VGG-neural net using Flipout as sampling weight method and calibrated after training).   

\begin{table}[h!]
  \centering
  \scalebox{0.66}{
\begin{tabular}{|l||l|l|l|l|l|l|}
\hline
\multicolumn{6}{|c|}{Statistics  for various MCMC sampling configurations}                                                                                                                                                                                                                                          \\ \hline
\multicolumn{1}{|l||}{\backslashbox{Metrics}{Map}} & \multicolumn{2}{l|}{\textbf{Temperature}}& \multicolumn{3}{l|}{\textbf{Temperature+Polarization}}  \\
\cline{2-6}\hline\rule{0pt}{10pt}
 &\multicolumn{1}{l|}{\textbf{MCMC}}& \multicolumn{1}{l|}{\textbf{covarBNN}} & \multicolumn{1}{l|}{\textbf{MCMC}} & \multicolumn{1}{l|}{\textbf{covarBNN}}& \multicolumn{1}{l|}{\textbf{Full-sky}} \\
\hline\rule{0pt}{10pt}
$ \omega_{b}$                                           & $0.0190^{+0.0013}_{-0.0013}$ &  $0.0190^{+0.0012}_{-0.0012}$ & $0.01967^{+0.00066}_{-0.00066}$ &  $0.01968^{+0.00064}_{-0.00064}$            &       $ 0.02009^{+0.00010}_{-0.00010}$     \\ \hline
\rule{0pt}{10pt}$\ln(10^{10}A_s)$                       &$3.633^{+0.031}_{-0.031}    $    &$3.633^{+0.031}_{-0.030}   $ & $3.648^{+0.015}_{-0.015}   $   &               $3.648^{+0.015}_{-0.016}     $&     $3.6449^{+0.0027}_{-0.0027}  $     \\ \hline \rule{0pt}{10pt}
$\omega_{cdm}$                                           & $0.171^{+0.011}_{-0.011}   $   &$0.170^{+0.011}_{-0.011}   $ & $0.1734^{+0.0031}_{-0.0032}   $   &          $0.1734^{+0.0031}_{-0.0031}   $   &   $0.1736^{+0.0009}_{-0.0009}$    \\ \hline\rule{0pt}{10pt}
$\Omega_{\Lambda}$                                        &  $0.583^{+0.025}_{-0.025}   $  &  $0.583^{+0.024}_{-0.025}   $ &  $0.5769^{+0.0079}_{-0.0080}   $   &           $0.5769^{+0.0079}_{-0.0079}   $&$ 0.5793^{+0.0019}_{-0.0019}$       \\ \hline\rule{0pt}{10pt}
Runtime                                                  &$4.02$hr &$1.56$hr &$4.40$hr &$3.14$hr &$4.52$hr//${\bf 3.15}$hr \\ \hline\rule{0pt}{10pt}
Acc. rate                                                  &$0.19$ &$0.23$ &$0.14$ &$0.25$ &$0.18$//${\bf 0.23}$\\ \hline\rule{0pt}{10pt}
$R-1$                                                  &$0.0093$ &$0.0098$ &$0.0051$ &$0.0084$ & $0.0091$//${\bf 0.0090}$\\ \hline\rule{0pt}{10pt}
$(R-1)_{95\%CL}$                                                  &$0.0827$ &$0.0764$ &$0.0944$ &$0.0642$ & $0.0940$//${\bf 0.0800}$\\ \hline

\end{tabular}}

\caption{\it Statistics and Parameters $95\%$ intervals for the minimal base-$\Lambda$CDM model from our synthetic CMB dataset using  a non-informative priori (MCMC) and a  precomputed covariance matrix from VI (covarBNN). The last column reports the metrics using the Full CMB map. The bold values in the last column correspond to the implementation of a  proposal posterior distribution from VI. Although the full sky gives the smallest credible region, MCMC is 10000 times slower than VI. The real value considered is the same as specified in Fig.~\ref{fig:mcmc}} \label{table:mcmctimes}
\end{table}
\begin{figure}[h!]
\begin{center}
\includegraphics[width=0.5\textwidth]{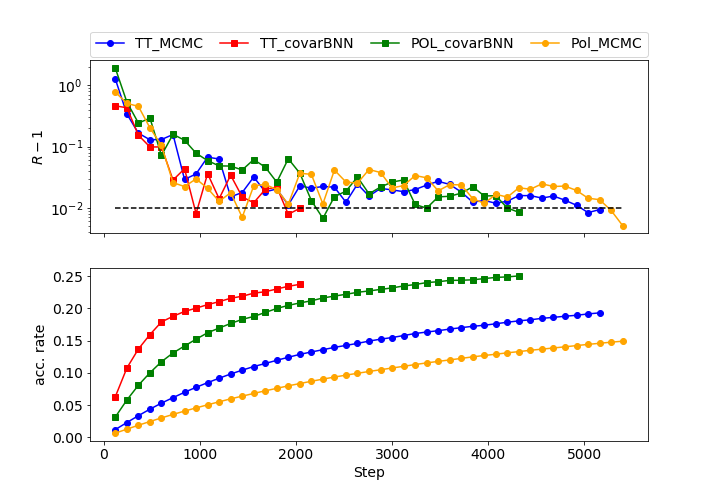}
\end{center}
\caption{\it  
Diagnostic information on the convergence of the MCMC
using unknown a priori (MCMC) and a precomputed covariance matrix from VI (covarBNN). (Top) Gelman-Rubin values
with respect to the acceptance step. The black dashed line $R-1=0.01$ stands for a necessary but not sufficient condition for stopping the run. (Bottom) Acceptance rate
with respect to the Metropolis-Hasting step.
Notice that using a proposal covariance matrix takes the MCMC reaches a higher acceptance rate in a lower number of steps. } \label{fig:mcmc_step}
\end{figure}
We  observe that  MCMC provides  tighter and more accurate constraints. However, the trained Neural Network can generate 8000 samples in approximately ten seconds which in turns out to be 10000 times faster  than MCMC for this dataset~\footnote{We run all single  MCMC experiments in  a CPU  Intel Core i7-3840QM with clock speed of 2.80GHz, while the BNN was trained in a GPU: GeForce GTX 1080 Ti.}. Runtime and metrics for convergence in  MCMC are shown in Table~\ref{table:mcmctimes}. As  expected, the polarization combined with temperature data  shifts the values obtained from temperature alone and  enhances the  accuracy in all parameters (columns 1 and 3). Furthermore, a well-converged chain  is also observed via the  Gelman-Rubin $R-1$ parameter and its standard deviation  at  $95\%$ confidence level interval $(R-1)_{95\%CL}$ (the smaller the better), so that the posterior  is accurately measured.
The qualitative correlations among parameters as obtained from BNNs are mostly analogous to the MCMC ones (Fig.~\ref{fig:mcmc}), showing that a multi-channel BNN is able to handle the complexities involved in this kind of analysis and additionally to use the polarization information to break cosmological degeneracies.
Nonetheless, although the MCMC is still 10000 times slower, it is able to better quantify the uncertainty. This is especially true when using the power spectrum of the full map, and the intervals are an order of magnitude more accurate than those computed by VI (rightmost column of Table~\ref{table:mcmctimes}). It would be  interesting (and more of a fair comparison) to compare MCMC for a full sky with respect to spherical neural architectures~\cite{2019arXiv190204083K,Perraudin:2018rbt} which can extract large scale signals correlations, thus  determining if Deep Learning methods can achieve a similar level of accuracy to MCMC.  On the other hand, we can also combine MCMC and VI leveraging the advantages of both methods. Such topic has attracted a lot of attention in the recent literature~\cite{10.5555/3045118.3045248,thin2020metflow}. A straightforward approach to speed up MCMC algorithms in big data problems consists in  using the covariance matrix constructed from the chains of the trained Neural Network as proposal for the distribution of the MCMC. In fact, it is known that a good estimate of the  covariance matrix for the parameters
increases the  acceptance rate leading to significantly faster convergence~\cite{PhysRevD.87.103529}.
In Table~\ref{table:mcmctimes}, we compare the runtime for the MCMC with and without a precomputed covariance obtained from BNN. As we can see from the table, proposal covariances from BNNs (covarBNN) speed up convergence in MCMC reducing the  computational time  for all datasets (Temperature, Polarization and full sky maps). In Fig.~\ref{fig:mcmc_step} we report MCMC convergence diagnostic quantities such as $R-1$ and the acceptance rate per iteration. The stopping  rule implemented in cobaya  ensures that the Gelman-Rubin $R-1$ value and its standard deviation  at  $95\%$ confidence level interval $(R-1)_{95\%CL}$ computed  from  different chains (four in our case), satisfy the convergence criterion $R-1<0.01$ twice in a row, and  $(R-1)_{95\%CL}<0.2$ respectively to stop the run~\cite{torrado2020cobaya}. For the Temperature signal alone, the Markov chains achieve a steady state in about 2000 steps working with the covarBNN proposal while it usually takes more than 5000 steps instead with the vanilla MCMC. This finding can also be explained by observing the acceptance rate in Fig.~\ref{fig:mcmc_step} (bottom), this value quickly approximates to around 0.23 allowing for a reasonably high acceptance probability (For more details see~\cite{roberts1997}). An analogous trend can be seen for the polarization case.  Motivated by this  discussion, the use of VI in cosmological dataset for either making inference directly in the maps or as a fast-method to speed up the MCMC techniques provides a promising tool in cosmology. The rest of this paper attempts to highlight the potential use and the critical questions with respect to BNN, and aims to serve as a guide for MCMC practitioners and cosmologists to the training and calibration of a BNN.

\section{\label{sectII} Epistemic and Aleatoric Uncertainty}
There are many  sources of uncertainty in model prediction of physical phenomena, and their nature   depend on the context and the application. However, these uncertainties have been categorized  in two groups: {\it aleatoric} and {\it epistemic} \cite{KIUREGHIAN2009105}.  Aleatoric uncertainties represent the intrinsic randomness in the input dataset \cite{Gal2016Uncertainty} and they can be reduced  enhancing the quality of the data. Moreover, this uncertainty can be {\it heteroscedastic}, i.e., the variability of the residuals does  vary as the independent variables do, or the uncertainty can be {\it homoscedastic} when it does not.

For any neural network, the  aleatoric uncertainty can be obtained by computing the variance of the  conditional distribution of the predictions given the features \cite{tagasovska2018singlemodel}. If such conditional distribution is Gaussian, the output of the network can be  split into mean predictions and their variance.  Then, the variance can be learned  implicitly from the minimization of the Gaussian log-likelihood while we supervise the learning of the regression task~\cite{NIPS2011_4329,Gal2016Uncertainty}.  Further methods to estimate and model the aleatoric uncertainty are given in~\cite{WANG201934,he2018analysis,PhysRevD.100.063514,tagasovska2018singlemodel}. \\ 
On the other hand, epistemic uncertainty quantifies the  ignorance about the correct model that generated the data, it  includes the uncertainty in the model, parameters, and convergence, among the others~\cite{KIUREGHIAN2009105}. This uncertainty is caused  by the limited training data with respect to the entire feature space.  Collecting more data in regions where there is a low density of training examples will reduce this uncertainty, while the aleatoric  will remain unchanged~\cite{Gal2016Uncertainty}.  Methods for estimating  epistemic uncertainties are different from the aleatoric ones, and this is where  BNNs can offer a mathematically grounded base for computing this  uncertainty and  be able to estimate the performance of the model~\cite{Gal2016Uncertainty}. Alternatives techniques for obtaining this uncertainty can be seen in~\cite{lakshminarayanan2016simple}.
Deep neural networks  involve both types of uncertainties and determining whether a particular uncertainty  is aleatoric  or epistemic but sometimes could be confused~\cite{KIUREGHIAN2009105}. However, for BNNs the authors in \cite{kwon,DBLP:journals/corr/abs-1806-05978,2018arXiv180605978S} rewrote the estimator for the variance such that it can be split in two terms associated to the epistemic and the aleatoric uncertainties. We will see in the Sec.~\ref{sectVII} that this split allows us  to evaluate the quality of the predictive uncertainty estimates. A more complete discussion of the nature of uncertainties can be found in~\cite{KIUREGHIAN2009105}.

\section{\label{sectIII}
Capturing Uncertainty In Neural Network Inference}
In this section we will briefly introduce some  variational inference techniques to deal with  non-tractable posterior distributions. We remind the reader to refer to~\cite{NIPS2011_4329,Gal2016Uncertainty,kwon} for additional details.\\
Let  $\Dataset=\{(\bm{x}_1, \bm{y}_1), \dots, (\bm{x}_D, \bm{y}_D)\}$ be a dataset formed by $D$ couples of inputs $\bm{x}_i \in \reals^M$  and their respective targets $\bm{y}_i \in \reals^N$, and $f_{W}({\bm x})$ be the output of the neural network with parameters (weights and biases) ${\bm w} \in \Omega$, where $\Omega$ is the parameter space.  
 
Neural networks are commonly trained by Maximum Likelihood Estimation (MLE), i.e., the parameters ${\bm w}$ are estimated in such a way that the likelihood of the observations in $\Dataset$ is maximized.
In the Bayesian setting, we choose a prior on the weights $p({W})$, and a model which allows the definition of the likelihood $p({\bm y}|{\bm x},{\bm w})$ capturing the predictive probability of the model given ${\bm w}$. The aim then is to find the posterior distribution given the observed dataset $p({\bm w}|\Dataset)$, which using Bayes' theorem can be written  as
\begin{equation}\label{eq:1}
  p({\bm w}|\mathcal{D})=\frac{p(\mathcal{D}|{\bm w})p({\bm w})}{p(\mathcal{D})}=\frac{\prod^D_{i=1} p({\bm y}_i|{\bm x}_i,{\bm w})p({\bm x}_i) p({\bm w})}{p(\mathcal{D})}
\end{equation}
where $p(\mathcal{D})=\int_\Omega p(\mathcal{D},{\bm w})d{\bm w}$ is the evidence, and the second equality holds assuming that $\mathcal{D}$ is a realization of i.i.d.~random variables~\cite{kwon}.
Once the posterior has been computed, the probability distribution of $\bm{y}^*$ for a new input $\bm{x}^*$ can be obtained by integrating out the parameters ${\bm w}$ as
\begin{equation}\label{eq:2}
  p({\bm y}^*|{\bm x}^*,\mathcal{D})=\int_{\Omega} p({\bm y}^*|{\bm x}^*,{\bm w}) p({\bm w}|\mathcal{D}) d{\bm w}.
\end{equation}
Unfortunately, the  posterior $p({\bm w}|\mathcal{D})$ usually cannot  be obtained analytically and thus approximate methods are commonly used to perform the inference task.
Here we will focus on a variational inference approach which approximates the posterior distribution $p({\bm w}|\mathcal{D})$ by an  variational distribution $q({\bm w}|\theta)$, chosen in  a well behaved functional space and depending on a set of variational parameters ${\bm \theta}$. The objective can then be formalized as finding ${\bm \theta}$ that makes $q$ as close as possible to the true posterior, for instance by minimizing the KullBack-Leibler (KL) divergence between the two distributions
\begin{equation}\label{eq:3}
\text{KL}(q({\bm w}|{\bm \theta})||p({\bm w}|\mathcal{D}))\equiv\int_\Omega q({\bm w}|{\bm \theta}) \ln \frac{q({\bm w}|{\bm \theta})}{p({\bm w}|\mathcal{D})}d{\bm w}.
\end{equation}
By substituting the true posterior given in Eq.~\eqref{eq:1} into Eq.~\eqref{eq:3}, we  can observe that minimizing the KL divergence   is equivalent to minimizing the following objective  function
\begin{equation}\label{eq:4}
\begin{split}
\mathcal{F}(\mathcal{D},{\bm \theta}) = \text{KL}&(q({\bm w}|{\bm \theta})||p({\bm w}))\\
-&\sum_{(\bm{x},\bm{y})\in\Dataset}\int_\Omega q({\bm w}|{\bm \theta}) \ln p({\bm y}|{\bm x},{\bm w}) d{\bm w},
\end{split}
\end{equation}
which is  often known as the variational free energy \cite{NIPS2011_4329}. The first term is the KL divergence between the variational distribution and the prior that acts as an Occam's razor term, i.e., it penalizes complexity priors, while the second term drives the variational distribution to place where the likelihood is high and the data is well explained~\cite{Gal2016Uncertainty}. Thus variational inference transforms  Bayesian learning from an analytically intractable integration to a manageable optimization problem. 

Suppose the objective function $\mathcal{F}(\mathcal{D},{\bm \theta})$ is minimized after the network is trained for some value ${\hat{{\bm \theta}}}$ of the variational parameters, then Eq.~\eqref{eq:2} can be rewritten in terms of $q({\bm w}|\hat{{\bm \theta}})$ as
\begin{equation}\label{eq:approxpost_pygivenx}
  q_{\hat{{\bm \theta}}}({\bm y}^*|{\bm x}^*)=\int_{\Omega} p({\bm y}^*|{\bm x}^*,{\bm w})q({\bm w}|{\hat{\bm\theta}})d{\bm w},
\end{equation}
where $q_{\hat{{\bm \theta}}}$ is the approximate predictive distribution . The authors in \cite{Gal2016Uncertainty} proposed an unbiased Monte-Carlo estimator for Eq.~\eqref{eq:approxpost_pygivenx}
\begin{equation}\label{eq:5}
 q_{\hat{{\bm \theta}}}({\bm y}^*|{\bm x}^*) \approx \frac{1}{K}\sum_{k=1}^K p({\bm y}^*|{\bm x}^*,\hat{{\bm w}}_k), \quad \mbox{with } \hat{{\bm w}}_k \sim q({\bm w}|{\hat{\bm\theta}}) \;,
\end{equation}
where $K$ is the number of samples.
We can also compute the covariance of the variational predictive distribution, for a fixed ${\bm x}^*$, by  invoking the total covariance law
\begin{align}\label{eq:6}
  \mathrm{Cov}&_{q_{\hat{{\bm \theta}}}}({\bm y}^*,{\bm y}^*|{\bm x}^*) \equiv \mathbb{E}_{q_{\hat{{\bm \theta}}}}[{\bm y}^*{\bm y}^{*\mathrm {T}}|{\bm x}^*] - \mathbb{E}_{q_{\hat{{\bm \theta}}}}[{\bm y}^*|{\bm x}^*]\mathbb{E}_{q_{\hat{{\bm \theta}}}}[{\bm y}^*|{\bm x}^*]^{\mathrm {T}}\nonumber\\
  &=\int_\Omega  \mathrm{Cov}_p({\bm y}^*,{\bm y}^*|{\bm x}^*) q({\bm w}|\hat{{\bm \theta}}) d{\bm w} \ + \nonumber \\
  & \quad + \int_\Omega \left[ \big(\mathbb{E}_p \Big[{\bm y}^*|{\bm x}^*]- \mathbb{E}_{q_{\hat{{\bm \theta}}}}[{\bm y}^*|{\bm x}^*]\big) \right. \times \nonumber \\
  &\quad \quad \quad \times \left. \big(\mathbb{E}_p[{\bm y}^*|{\bm x}^*]- \mathbb{E}_{q_{\hat{{\bm \theta}}}}[{\bm y}^*|{\bm x}^*]\big)^{\mathrm {T}}\right]q({\bm w}|\hat{{\bm \theta}}) d{\bm w},
\end{align}
where $\mathbb{E}_{q_{\hat{{\bm \theta}}}}[{\bm y}|{\bm x}] = \int {\bm y}\, q_{\hat{{\bm \theta}}}({\bm y}|{\bm x}) d{\bm y}$,
${\bm y}^{\mathrm{T}}$ is the transpose of the vector $\bm {y}$, and $\mathbb{E}_p[{\bm y}|{\bm x}]=\int {\bm y}\, p({\bm y}|{\bm x},{\bm w}) d{\bm y}$ \cite{kwon,2018arXiv180605978S}. 
The first term in Eq.~\eqref{eq:6} collects the variability of the output coming from the training dataset which corresponds to the aleatoric uncertainty as it was mentioned in the previous section, while the second term encodes the variability of the output coming from the model, which it should be associated to the epistemic uncertainty. Following \cite{kendall2017uncertainties,Cobb_2019,Dorta_2018},  we assume that the last layer of the network  consists of a mean vector $\bm{\mu}\in\mathbb{R}^{N}$ and a covariance matrix $\Sigma\in\mathbb{R}^{N(N+1)/2}$. Suppose that for a given fixed input $\bm{x}^*$, $T$ forward passes of the network are computed, obtaining for each of them a mean vector $\bm{\mu}_t$ and a covariance matrix $\Sigma_t$. 
Then, an estimator for Eq.~\eqref{eq:6} can be written as  
\begin{equation}\label{eq:7}
\widehat{\mathrm{Cov}}(\bm{y}^*,\bm{y}^*|\bm{x}^*)\approx \underbrace{\frac{1}{T}\sum_{t=1}^{T}\Sigma_t}_\text{Aleatoric}+ \underbrace{\frac{1}{T}\sum_{t=1}^{T}( {{\bm{\mu}}}_{t}-\bm{\overline{\mu}})( {\bm{\mu}}_{t}-\bm{\overline{\mu}})^{\mathrm{T}}}_\text{Epistemic}, 
\end{equation}
with $\bm{\overline{\mu}}= \frac{1}{T}\sum_{t=1}^{T} {\bm{\mu}}_t$.
 Notice that in case $\Sigma$ is dia\-gonal, with $\bm{\sigma}^2 = \text{diag}(\Sigma)$, the last equation reduces to the variance of the variational predictive distribution given in~\cite{kendall2017uncertainties,kwon}, given by the sum of both the aleatoric and the epistemic uncertainties, that is 
\begin{equation}\label{eq:8}
\widehat{\mathrm{Var}}( \bm y^*|\bm{x}^*)\approx \underbrace{\frac{1}{T}\sum_{t=1}^{T}\bm \sigma^2_{t}}_\text{Aleatoric}+ \underbrace{\frac{1}{T}\sum_{t=1}^{T}{\bm \mu}^2_{t}-\bm{\bar{\mu}}^2}_\text{Epistemic}. 
\end{equation}
In this setting, neural
network can be used to learn the correlations between the the targets and
produce estimates of their uncertainties.

\section{\label{sectIV} Variational Distributions}
\begin{figure}[h!]
\begin{center}
\includegraphics[width=0.48\textwidth]{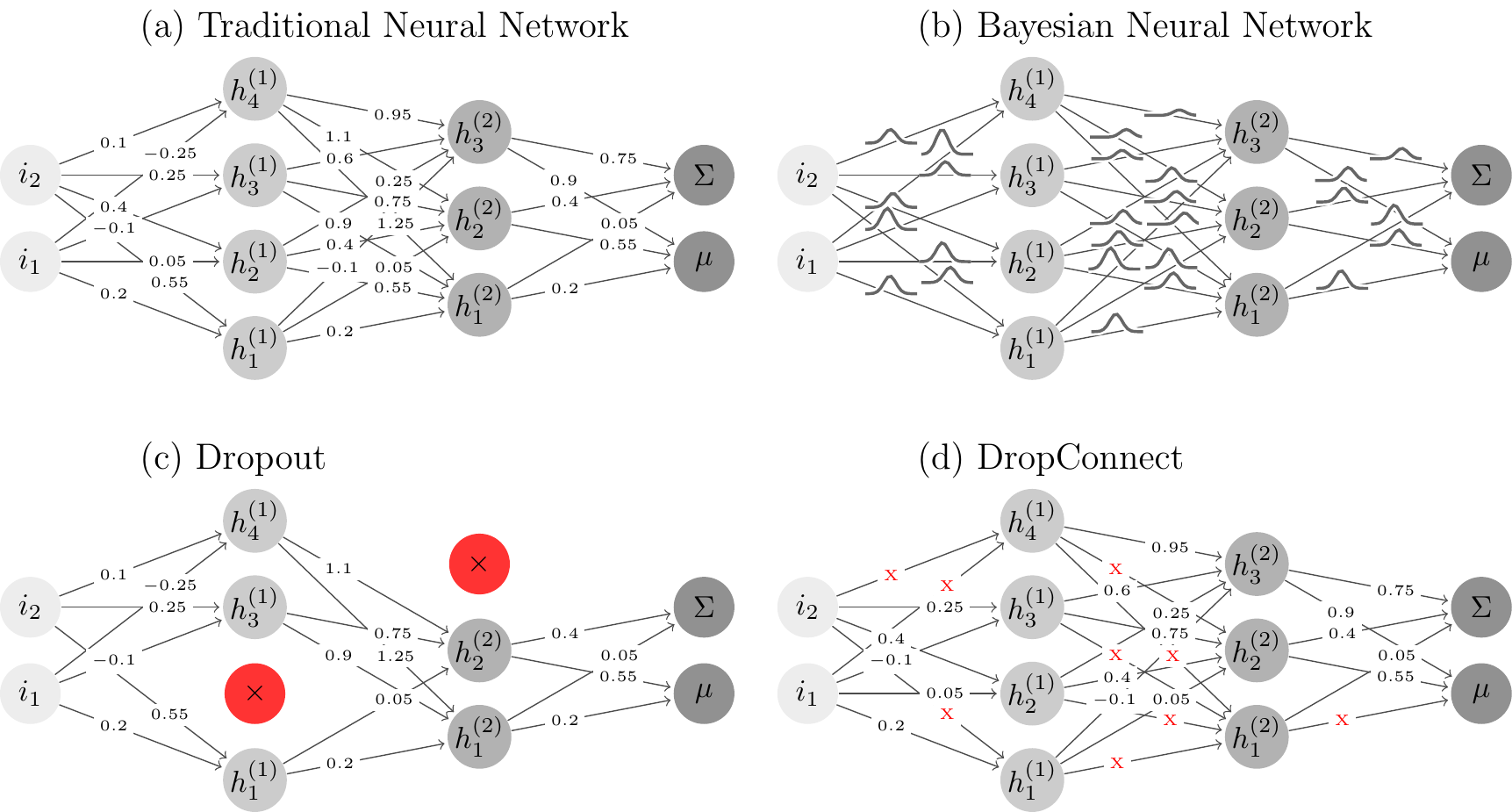}
\end{center}
\caption{\it  Diagrammatic representation of aleatoric neural networks with a stochastic Gaussian layer in output. The nodes $i$ correspond to the inputs of the network (e.g., the pixels of an image); the nodes $h$ are the nodes in the hidden layers; the output of the networks are the mean and covariance of a Gaussian distribution.} \label{fig:BNN}
\end{figure}

In this section we will review different types of neural networks, all characterized by a common (aleatoric) Gaussian layer  in output. After training, in deterministic neural networks (Fig.~\ref{fig:BNN}a) the weights have a fixed value. On the other side, in Bayesian Neural Networks (Fig.~\ref{fig:BNN}b) we have a prior and a posterior distribution defined over their weights (Sec.~\ref{sectIII}), which is usually chosen to belong to a well behaved family of distributions. Two popular approximations for BNNs are Dropout and DropConnect. In Dropout (Fig.~\ref{fig:BNN}c) each neuron is dropped with a certain probability, while in DropConnect (Fig.~\ref{fig:BNN}d) the weight-connections are dropped instead. The most popular approaches for BNNs present in the literature are briefly summarized in the following.

\subsection{Bernoulli via Dropout}
Dropout was first proposed in \cite{JMLR:v15:srivastava14a} as a regularisation method for neural networks which helps to reduce co-adaptations amongst the neurons. 
During training, each neuron in the $j$-th layer $\bm{h}^{(j)}$ of size $H_j$ is dropped from the network with probability $p$ (commonly known as Dropout rate). The application of Dropout  can be expressed as
\begin{equation}\label{eq:9}
  \bm{h}^{(j+1)}=\sigma \left(\bm{m}^{(j+1)}\circ ({W}^{(j)}\, \bm{h}^{(j)}) \right), \ \bm{m}^{(j+1)}\sim \text{Ber}(p),
\end{equation}
where $\circ$ corresponds to the Hadamard product, $\sigma(\cdot)$ is a nonlinear activation function, ${W}^{(j)}$ is the weight $(H_{j+1} \times H_{j})$-matrix for the layer $j$, and $\bm{m}^{(j+1)}$ a vectorial mask of size $H_{j+1}$, which is sampled from a Bernoulli distribution with probability $p$.
Once trained, the entire network is used  although  neurons  are scaled using the factor $1-p$, this compensates for the larger size of the network compared to the one used during training.
Interestingly, the authors in \cite{pmlr-v48-gal16}  have shown a connection between Dropout and approximate variational  inference  for Gaussian processes, allowing the  neural network to be interpreted as an approximate Bayesian model.  In this case, the variational distribution $q({W}^{(j)})$ for the $j$-th layer associated to Eq.~\eqref{eq:9} can be defined as~\cite{Gal2016Bayesian} 
\begin{equation}\label{eq:10}
  {W}^{(j)}={V}^{(j)} \,  \text{diag} \left({\bm{m}}^{(j)} \right), \quad {\bm{m}}^{(j)}\sim \text{Ber}(p),
\end{equation}

being ${V}^{(j)}$ a $(H_{j+1}\times H_{j})$-matrix of variational para\-meters to be optimised. Inserting this variational distribution into Eq.~\eqref{eq:4}, we obtain an unbiased estimator for the objective  function~\cite{pmlr-v48-gal16}
\begin{equation}\label{eq:11}
-\hat{\mathcal{F}}= \sum_{i=1}^D \ln p({\bm y}_i|{\bm x}_i,{\bm w})-\lambda\sum_{j=1}^L|| {W}^{(j)}||^2,
\end{equation}
 where $\lambda$ is a positive constant and the weights are sampled at each layer from  $q({W}^{(j)})$ defined in Eq.~\eqref{eq:10}. The first term corresponds to the likelihood that encourages ${\bm w}$ to  explain well the  observed  data, while the second term is a $L_2$ regularization, weighted by the weight decay parameter $\lambda$, which mimics the KL divergence term in Eq.~\eqref{eq:4}. Therefore,  training  a neural  network using Dropout has  the  same  effect  as  minimizing  the KL divergence in Eq.~\eqref{eq:3}. This scheme, besides  working similar to a Bayesian Neural Network, acts also as a  regularization method which prevents over-fitting. After training the neural network, Dropout remains active  and we  follow  Eqs.~\eqref{eq:5} and \eqref{eq:7}  to  perform inference and estimate the uncertainties of the network. Such procedure is known in the literature as Monte Carlo Dropout. 

\subsection{Bernoulli via DropConnect} 
DropConnect is a  generalization of Dropout used for regularization in deep neural networks~\cite{pmlr-v28-wan13}. In this method, each weight-connection is dropped  with  probability $p$, 
differently from DropConnect where instead neurons are dropped.
The hidden nodes of each layer are given by 
\begin{equation}\label{eq:12}
  \bm{h}^{(j+1)}=\sigma \left(({M}^{(j)} \circ {W}^{(j)})\, \bm{h}^{(j)} \right), \quad {M}^{(j)}\sim \text{Ber}(p)~,
\end{equation}
where ${M}^{(j)}$ a matrix mask of size $H_{j+1} \times H_{j}$. In~\cite{2019arXiv190604569M,McClure2016RepresentingIU} the authors use DropConnect to obtain approximated uncertainties. Here, the mask  is  applied  directly  to  each weight, differently from Dropout where the weights are not sampled. The variational distribution for the weights of the $j$-th layer is defined by
\begin{equation}\label{eq:13}
  {W}^{(j)}={W}^{(j)}\circ {{M}^{(j)}}, \quad {M}^{(j)}\sim\text{Ber}(p),
\end{equation}
where $m^{{(j)}}_{rs}$ is the element of the mask ${M}^{(j)}$ connecting  the $r$-th neuron  of  the $(j+1)$-th layer  to the $s$-th neuron  of  the $j$-th layer. Similarly to the previous case, during training  the network weights are learned to minimize Eq.~\eqref{eq:11}, while at test time each input is passed through the network multiple times.  DropConnect allows to capture  both the epistemic and aleatoric uncertainties via Eq.~\eqref{eq:7} . 

\subsection{Reparameterization Trick and Flipout}
So far we have seen two methods to provide the network with stochastic weights. Dropout deals with stochastic activations (drop neurons), the weights are not sampled independently, however it is easy to implement and quite cheap to compute. On the other hand, DropConnect drops directly the weights which in most cases are far more than the number of neurons, i.e., this method is more expensive and it has a higher variance. 

Recently, different works have proposed to sample the weights from a Gaussian distributions instead. The {\it Reparameterization Trick} (RT) allows to generate  samples which are  differentiable with respect to the the parameters of the distribution from which they are drawn \cite{2013arXiv1312.6114K}. If the weights are  considered as a continuous random variable drawn from $\bm{w}\sim q({\bm w}|{\bm \theta})$, thanks to the RT we might  express it  as a deterministic function  $\bm{w}= g(\bm{\epsilon}, {\bm \theta})$ of a fixed random auxiliary variable $\bm{\epsilon}$, i.e., $\bm{\epsilon}\sim p(\bm{\epsilon})$ has a probability density  function $p$ independent of $\theta$, while $g$ is parameterized by $ {\bm \theta}$. This implies that any expectation with respect to $q({\bm w}|{\bm \theta})$, can be estimated as \cite{2013arXiv1312.6114K,2015arXiv150602557K}
\begin{equation}\label{eq:14}
\int_\Omega q({\bm w}|{\bm \theta})f({\bm w})d{\bm w}\approx \frac{1}{K}\sum_{k=1}^K f( g(\bm{\epsilon}_k, {\bm \theta})) \;\; \mbox{with } \bm{\epsilon}_k \sim p(\bm{\epsilon}).
\end{equation}
In the multivariate Gaussian case $\bm{w}\sim \mathcal{N}({\bm{\mu}},{\Sigma})$ \cite{2017arXiv171200424W}, we have  the usual reparameterization  given by $\bm{w}=\bm{\mu}+{ L}\,\bm{\epsilon}$, where $\bm{\epsilon}\sim \mathcal{N}(\bm{0},{I})$ and ${L}$ has the property that  ${\Sigma}={LL}^\top$, a noteworthy example being the lower triangular Cholesky factorization. Given this reparameterization along with Eq.~\eqref{eq:14}, we can get the approximated value of the second term in Eq.~\eqref{eq:4} and thus it is possible to derive the unbiased  estimate of the gradient of the variational free energy \cite{2013arXiv1312.6114K}.

The downside of the RT is that the sampled weights are the same for all the examples in the batch, thus correlating the gradients between different samples in the same batch. To overcome this limitation and thus reduce the gradient variance, the authors of \cite{wen2018flipout} propose the {\it Flipout} method as an  efficient way to provide pseudo-independent weights perturbations. Methods like Flipout or Local Reparameterization Trick \cite{2015arXiv150602557K} are some of the strategies used today for variance reduction. Flipout assumes that the variational distribution can be written as a mean ${\overline{W}^{(j)}}$ plus a perturbation $\Delta W^{(j)}$ with symmetric distribution around zero, i.e., 
\begin{equation}
    {W}^{(j)} = {\overline{W}^{(j)}}+ {\Delta W^{(j)}} ~,
\end{equation}
and proposes to decorrelate the noise inside a mini-batch by sampling a series of pseudo-random sign matrices, to randomly flip the symmetric perturbation of the weights. For a sample $i$ in the batch
\begin{equation}
    {\Delta W}^{(j)}_i =  {\widehat{\Delta W}^{(j)}} \circ (\bm{r}_i\, \bm{s}_i^\top )\;
\end{equation}
where ${\bm r}_i$ and ${\bm s}_i$ are random vectors whose entries are uniformly sampled from $\{\pm 1\}$ and ${\widehat{\Delta W}}^{(j)}$ is a perturbation sampled only once for the whole mini-batch. Remarkably this approach can be easily vectorized for a given batch and used to efficiently obtain pseudo-random weights perturbations~\cite{2015arXiv150602557K}.

\section{\label{sectV} Dataset and Network} 

\begin{figure}[h!]
\begin{center}
\includegraphics[width=0.47\textwidth]{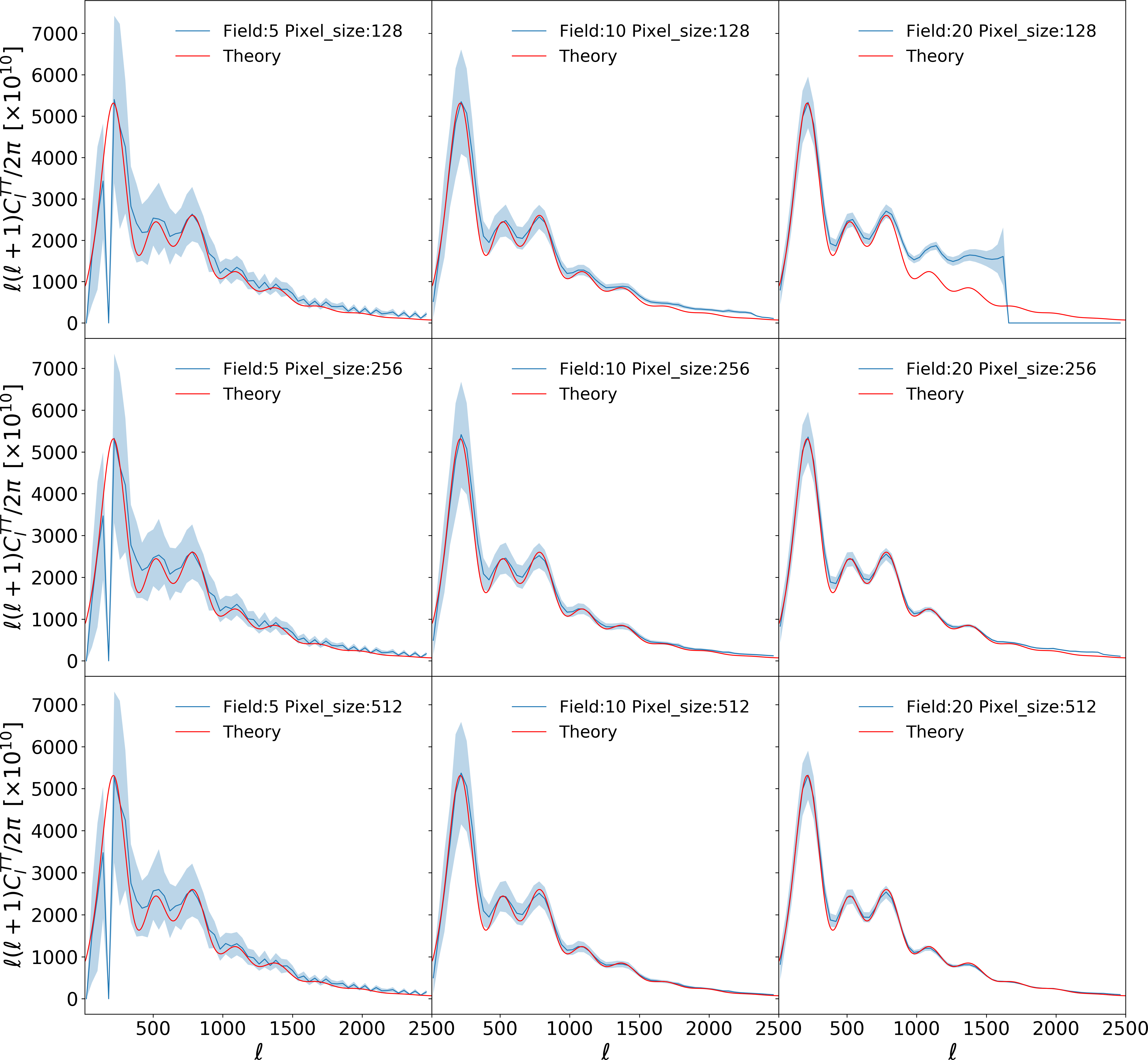}
\end{center}
\caption{\it Power spectra generated from patches with diffe\-rent angular scales (Field: \{5,10,20\} $deg^2$)  and pixelized into: \{128,256,512\} pixels).
The light-blue shadowed regions correspond to the standard deviation of 500 samples. The red line shows the power spectrum obtained from the theory. We have used lens-tools (using azimuthal averaging in Fourier space)  to compute the power spectrum of the CMB patches~\cite{2016A&C....17...73P}. The configuration used in the following section is 20$deg^2$ with 256 pixels.} \label{fig:1}
\end{figure}

We have generated 50,000 independent realizations of simulated CMB full-sky maps and extracted from them images of
20$\times$20 deg$^2$ and $256\times256$ pixels. From the total dataset, $70\%$ is reserved for training, $10\%$  for validation, and $20\%$  for testing.
These simulations have been created given the temperature angular power spectra generated by CLASS and healpy~\cite{Zonca2019}, by originating a realisation on the HEALPix grid.
The choice of the resolution for the maps comes from the analysis displayed in Fig.~\ref{fig:1}. Here we can see that for small angular sizes, the power spectrum obtained from the patches cannot retain enough information from the original  spectrum produced  by CLASS, in contrast to images with angular size equal to or larger than 10$\times$10 deg$^2$. However for  very larger angular size, the assumption of a good flat approximation is not valid and distortions produced by the projection of the spherical data into the flat sky could lead to undesirable effects, that is,  the resolution leads to a  trade-off between accuracy of the recovered power spectrum and execution time (e.g., speed) of the neural network. One alternative to deal with large angular size images is to create CMB maps from the lens-tools package~\cite{2016A&C....17...73P} which produces a Gaussian random field directly over the pixels of the flat image.

In this paper we assume a minimal version of the $\Lambda$CDM model where each power spectrum generated in CLASS differs in three parameters:  {\it baryon density} $\omega_b \in[0.019,0.031]$, {\it cold dark matter density} $\omega_{cdm} \in[0.06,0.22]$, and  {\it primordial spectrum amplitude} $A_s \in[1.01\times10^{-9},4.01\times10^{-9}]$, sampled over a uniform 3D grid, while  the rest of the $\Lambda$CDM  parameters are fixed to the values reported by the Planck mission~\cite{2018arXiv180706211P}. The multiple generations of the power spectra should gather the cosmic variance that will contribute to the aleatoric uncertainty. The images and the parameters are normalized between -1 and 1, without any additional data augmentation.

\subsection*{Architecture}
\begin{figure}[h!]
\begin{center}
\includegraphics[width=0.5\textwidth]{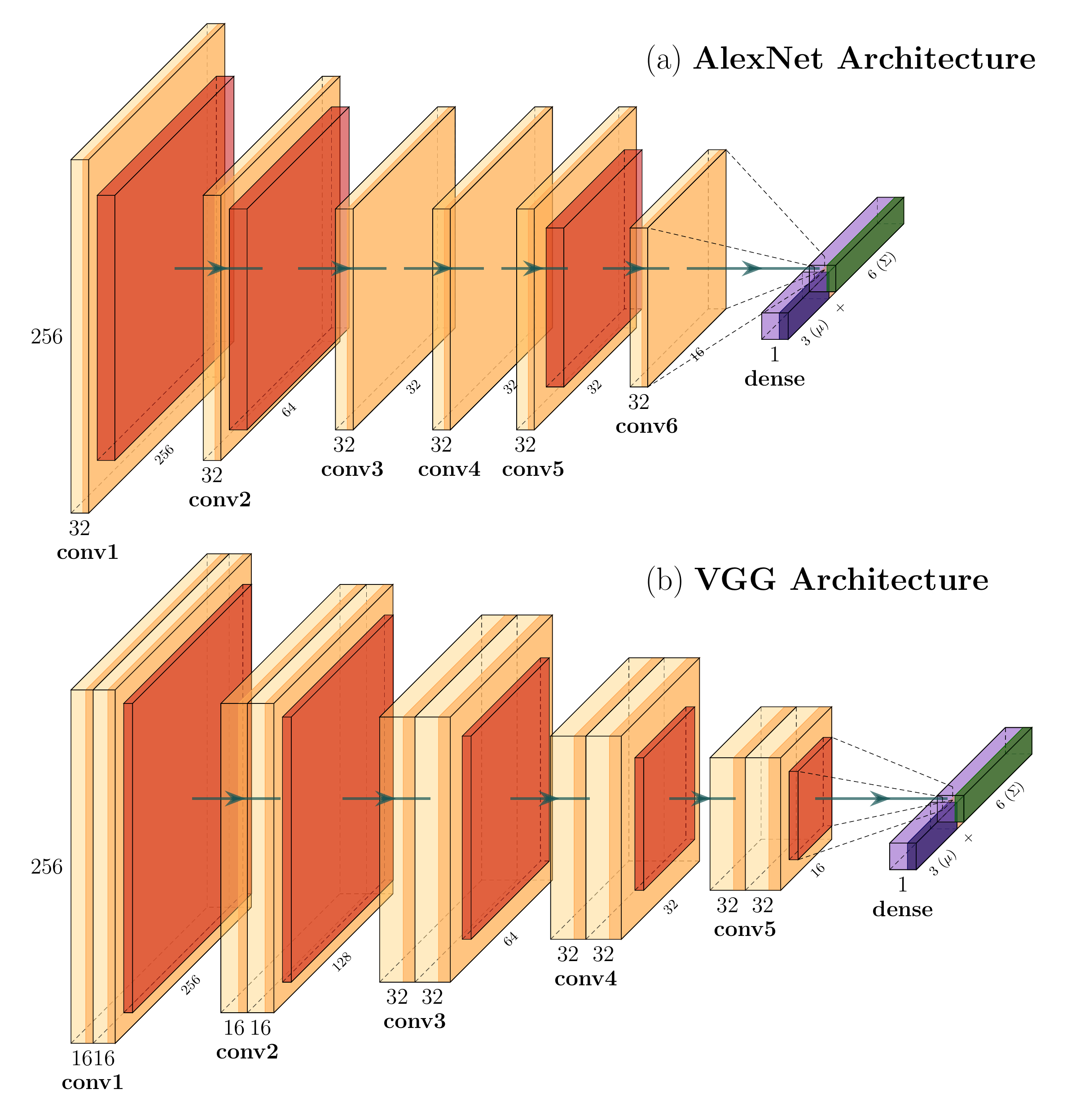}
\end{center}
\caption{\it  Illustration for the modified versions of the (a) AlexNet and (b) VGG architectures. The input of the networks are images of $256\times256$ pixels and the output consists of nine values. Both architectures have around $\sim60,000$ weights, being  batch normalization in VGG  the most remarkable difference between them.} \label{fig:vgg16}
\end{figure}

 We have implemented our models in TensorFlow~\footnote{\MakeLowercase{h}ttps://www.tensorflow.org/}. The API  tf.Keras and the library TensorFlow-Probability~\footnote{\MakeLowercase{h}ttps://www.tensorflow.org/probability} have been also used for RT and Flipout, while Dropout and DropConnect were implemented in the higher-level library Sonnet~\footnote{\MakeLowercase{h}ttps://sonnet.readthedocs.io/en/latest/}. 
 We  implemented  a  modified  version  of the VGG~\cite{7486599} and AlexNet~\cite{Krizhevsky:2012:ICD:2999134.2999257} networks  illustrated  in Fig.~\ref{fig:vgg16}.
We have chosen to have all the  architectures with  roughly the same number of weights so that the analysis carried out for all BNNs  depends only on their performance, and not on the size or complexity of the network.     
The  VGG network consists  of ten convolutional  layers with a fixed  kernel size of $3\times 3$,  using  LeakyReLU as the activation function.  Each convolutional layer, except for the last one, is followed by a batch renormalization layer, which ensures that the activations computed in the forward pass during training depend only on a single example  and are identical to the activations  computed in test \cite{Ioffe2017BatchRT}. We have applied zero  padding in each convolution layer, and we have downsampled  using max pooling, allowing the network to learn correlations at large angular scales. For AlexNet the input is  convolved with six convolutional  layers of kernel size  $(11,5,3,3,3,1)$,  with the same activation function after each layer, and without batch (re)normalization. The  downsampling is done using max pooling for three of the six layers as we can see in  Fig.~\ref{fig:vgg16}. One critical modification with respect to AlexNet consists in the change of the fully connected layers at the end of the network which are replaced by convolutional layers.  Indeed, we have observed that in our configuration, for the CMB dataset, the presence of the dense layers deteriorates the performances. At the end of the convolutional part for both architectures, a  dense layer with  nine neurons is built, three of them correspond to the means  of the cosmological parameters used to generate the maps, and the other six  compose a lower triangular matrix ${L}$ \cite{Dorta_2018,Cobb_2019}.
This last layer yields to a multivariate Gaussian distribution with mean $\bm{\mu}$ and covariance  $\Sigma={L}{L}^\top$ to guarantee positive definiteness.

\subsection*{Training}
The  negative log-likelihood (NLL) of our neural network, used to estimate the cosmological parameters and their uncertainties, is given by 
\begin{equation}\label{eq:16}
\mathcal{L}\sim \frac{1}{2}\log |\Sigma|+ \frac{1}{2}(\bm{y}-\bm{\mu})^\top \Sigma^{-1}(\bm{y}-\bm{\mu}),
\end{equation}
averaged over the mini-batch. The objective function differs depending on the method employed. For example, for Dropout and DropConnect, it will be expressed as the sum of the  negative log-likelihood in Eq.~\eqref{eq:16} plus a L2-regularization term from Eq.~\eqref{eq:11}. Here we used $\lambda=0.001$ for the weights and $\lambda=0.0001$ for the biases. 

In the case of RT and Flipout, the optimization is based on the minimization of the KL divergence written in Eq.~\eqref{eq:4}. The prior that we have chosen is a normal distribution under the mean field approximation, initialized with mean 0 and variance 1, while the posterior is given by another Gaussian distribution  initialized with the Glorot normal initializer for the mean, while the variance is sampled from a Gaussian  distribution $\mathcal{N}(-9,0.01)$ (before applying the Softplus function). Furthermore, the weights  of the posterior are controlled by a L2-regularization term and the biases in both cases are taken as a deterministic function. Different  experiments using  (non)-trainable prior distributions showed  that training both the posterior and the prior parameters  turned out in better performances. Furthermore, a deterministic dense layer as the last layer of the network (producing in output $\bm{\mu}$ and $\Sigma$) instead of a probabilistic one produces better results.  The algorithm used to minimize the objective function is the Adam optimizer  \cite{2014arXiv1412.6980K} with first and second moment exponential decay rates of 0.9 and 0.999, respectively, a learning rate of   $10^{-4}$,  and decay rate of $0.9$. The decay step has been tuned based on the specific method: for Flipout is  6,000, for Dropout and DropConnect is 8,000, and for RT 2,000. We trained the networks for 400 epochs with batches of 32 samples.

\subsection*{Validation and Test}
We have fed each input image from the test set 2,500 times to each network,  essentially  getting  enough samples from the posterior of the network and hence being able to capture the epistemic uncertainty.  Each sample produces nine variables corresponding to the cosmological parameters and their covariance matrix. The latter  represents the aleatoric uncertainty learned from optimizing the objective function, hence, the total uncertainty reported for each example is provided via Eq.~\eqref{eq:7}.

\section{Calibration}\label{sec:calibration}
The issue of the calibration  of  neural  networks  has gained  interest  in  the  recent  years,  since it has been shown  that deep neural networks  tend to be overconfident in their predictions~\cite{2019arXiv190602530O}. Different works have addressed to problem of identify why a neural  network may become miscalibrated (see, e.g.,~\cite{Guo:2017:CMN:3305381.3305518}, and references therein). One of the ways to diagnostic the quality of the  uncertainty estimates is  through   reliability diagrams. 
\begin{figure}[h!]
\begin{center}
\includegraphics[width=0.5\textwidth]{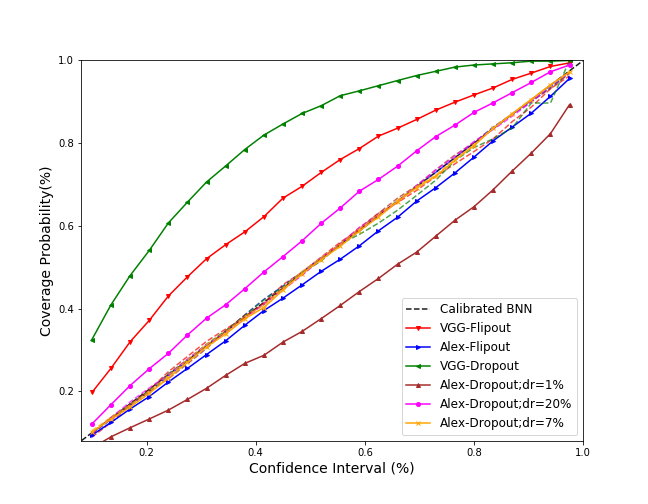}
\end{center}
\caption{\it Reliability diagrams for CMB maps before (solid lines with marks) and after calibration (dashed lines). The black dashed line stands for the perfect calibration, while the other color lines represent different BNNs. For some  hyperparameters values in AlexNet (Alex), the model underestimates its errors, while for VGG all models are overconfident in their predictions. We have implemented beta calibration to obtain these curves \cite{pmlr-v54-kull17a}.} \label{fig:calibration1}
\end{figure}
Fig.~\ref{fig:calibration1} displays the confidence intervals against the expected  coverage probabilities defined as the  $x\%$  of samples for which the true value of the parameters falls in the  $x\%$-confidence interval. If the network is well calibrated, then the diagram  should correspond to a straight line corresponding to the identity function and any deviation from it represents a miscalibration. As we will show later, the  methods employed to adjust the predicted uncertainties can be applied during or after training. During the training process, we just need to adapt some hyper parameters in the model in order to achieve a good calibration. For example, in the case of Dropout the authors in~\cite{PerreaultLevasseur:2017ltk} found out that the Dropout rate should be tuned to produce high accuracy uncertainty estimations (see AlexNet with Dropout rate $7\%$ in Fig.~\ref{fig:calibration1}).  Moreover, the authors in~\cite{2017arXiv170507832G}  introduced Concrete Dropout  which allows for the dropout probabilities to be  automatically   tuned, improving the performance and producing calibrated uncertainties. Additionally, we will show in the next section that the hyper parameters related to Flipout correspond to the regularization parameters for the scale of the approximate posterior over weights and biases. However, calibrating the network during training is not efficient in all cases. Tuning  the hyper parameters  could drastically affect the performance of the model and besides this, the method depends strongly on the architecture of the network. An example on this issue  will be shown later when we will observe that this technique fails on the VGG architecture.  On the other hand,  it has been noticed that methods for calibrating the network after training  indeed preserve the accuracy of the predictions  achieved during training.   Histogram Binning~\cite{Zadrozny:2001:OCP:645530.655658},  Isotonic Regression~\cite{Zadrozny:2002:TCS:775047.775151}, Platt Scaling~\cite{Platt99probabilisticoutputs,2018arXiv180700263K,pmlr-v54-kull17a}, and  Temperature Scaling are some of the most common methods used for calibrating the networks for regression tasks. In this work (except in Subsec.~\ref{secpolcal}) we will use an extended version of the parametric Platt Scaling method described in~\cite{pmlr-v54-kull17a}. Basically, we build the reliability diagram and fit it to the calibrated map~\cite{pmlr-v54-kull17a}
\begin{equation}\label{eq:17cal}
\beta(x;a,b,c)=\frac{1}{1+\big(e^{c}\, \frac{x^a}{(1-x)^b}\big)^{-1}},
\end{equation}
with  scalar parameters $a$, $b$, and $c$ $\in \mathbb{R}$. Hence, we apply the following transformation to the covariance matrix $\Sigma\rightarrow s\Sigma$ in the evaluation of the coverage probabilities, see Eq.~\eqref{eq:18cal}, being $s\in\mathbb{R}^+$ a scalar parameter. Finally, we choose the value of $s$ used for the calibration of the  network by minimizing the  difference of the calibrated maps with respect to  the diagonal line. Fig.~\ref{fig:calibration1} displays the results of the beta calibration for different BNN models for VGG and AlexNet. For example, a Dropout rate of $1\%$ using AlexNet or  Flipout on AlexNet without regularization on their posterior weights will produce underestimation in their errors. This means that most of the true values do not fall in their corresponding confidence intervals, as we can see in Fig.~\ref{fig:calibration2}. On the other hand, overconfident networks (like VGG, as we see in Fig.~\ref{fig:calibration1}) are very conservative in their errors, therefore they produce weak  confidence bounds on the parameter space, as it is shown in Fig.~\ref{fig:calibration2}.
\begin{figure}[h!]
\begin{center}
\includegraphics[width=0.5\textwidth]{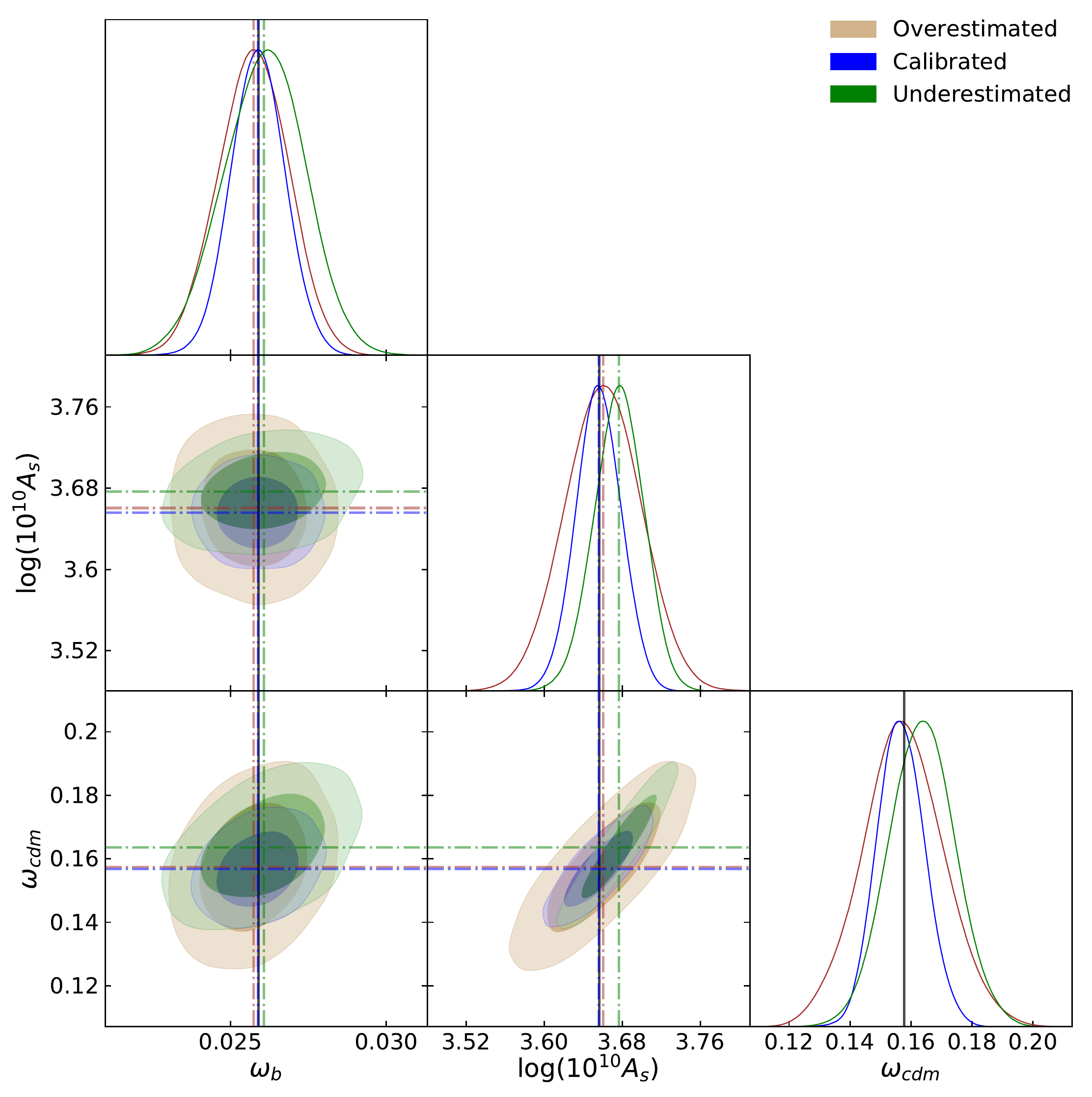}
\end{center}
\caption{\it Triangular plot for the $95\%$  contours of the cosmological parameters obtained from  calibrated (blue), under (green)- and over (orange)- estimated  models. The black lines stand for the real value, while the dashed lines refer to their predicted values. The marginalized distributions show the behavior of miscalibrated networks compared with a calibrated one. We have used the Python GetDist package for creating the triangular plot~\cite{lewis2019getdist}.} \label{fig:calibration2}
\end{figure}
Finally, let us discuss about the evaluation of the coverage probabilities. We have noticed how all models used in this work produce a full predictive posterior distribution (3) which is approximately Gaussian (higher order statistical moments very close to zero). Hence, treating the distribution as a multivariate Gaussian, the confidence intervals can becomputed by
\begin{equation}\label{eq:18cal}
\mathcal{C}\geq(\bm{y}-\bm{\hat{y}})^\top \Sigma^{-1}(\bm{y}-\bm{\hat{y}}),
\end{equation}
 which is basically an ellipsoidal confidence set with cove\-rage probability $1-\alpha$. The quantity $\mathcal{C}$ has the Hotelling's T-squared distribution  $T^2_{k,D-k}(1-\alpha)/D$, with $k$ degrees of freedom, being $D$ the number of samples~\cite{bimj.4710240520}. For large samples, the Hotelling's $T^2$ tends to the more common $\chi^2$ distribution~\cite{hotelling1931}, which is the distribution of the sum of the squares of $k$ independent standard normal random variables. This is indeed the distribution which we will use in the calculation of our confidence intervals. Therefore, the coverage probabilities correspond to the percentage of samples satisfying Eq.~\eqref{eq:18cal}, in other words, the fraction of  examples where the true values lay into the 3D-ellipsoidal region. This evaluation generalizes the methods in the literature in which we must bin the samples in order to estimate the region that contains $(1-\alpha)$ of the test dataset, as long as the joint distribution is almost Gaussian. In Appendix~\ref{appendixA1}, we compute the coverage probabilities from the histograms as it is usually done in the literature~\cite{Lanusse:2017vha}, finding consistent results when we used  ellipsoidal confidence intervals.

 \section{\label{sectVII} Analysis and Results}
 
In this section we describe the results we obtained with different architectures and types of BNNs.  We compare all experiments in terms of  performance, i.e.,  the precision  of  their  predictions  for  the  cosmological  parameters quantified  through Mean Square Error (MSE), and their values achieved in the  NLL. Furthermore, we  analyze the quality of the uncertainty estimates in each experiment and its appropriate calibration, if needed. 

\subsection{ Dropout and DropConnect}
We begin by comparing the performance of Dropout and DropConnect. The best results are displayed in Table~\ref{table:1}.  As it can be seen,  DropConnect does not exhibit particularly exciting performances for any architecture. Even for a vast range of regularization  and initialization  values, we could not achieve good  results, as was reported in~\cite{2019arXiv190604569M}.  It seems that DropConnect injects large noise on convolutional layers  until this unstabilizes the  training  process.   Nonetheless, in contrast to the DropConnect technique, Dropout  effectively   improves  the  performance of  VGG and AlexNet  by  a  noticeable  margin.
\begin{table}[h!]
 \scalebox{0.9}{
\begin{tabular}{|l||l|l|l|l|l|l|l|l|}
\hline
\multicolumn{9}{|c|}{Performance for different BNNs}                                                                                                                                                                                \\ \hline
\multicolumn{1}{|c||}{\backslashbox{}{model}} & \multicolumn{2}{c|}{\textbf{Dropout}}& \multicolumn{2}{c|}{\textbf{DropConnect}} & \multicolumn{2}{c|}{\textbf{RT}} & \multicolumn{2}{c|}{\textbf{Flipout}}  \\ \cline{2-9} 
\multicolumn{1}{|l||}{metrics}                                                                               & VGG                 & Alex  & VGG                 & Alex                & VGG                & Alex                & VGG                & Alex                          \\ \hline
MSE                                                                                            & 0.05               &0.1      & 0.45               &0.68             &0.08              & 0.30             & 0.04             &0.05                              \\ \hline
NLL                                                                                            & -3.17        &-1.39     &-0.13           &-0.08                   &-2.33                   &-2.01                   &-3.20                   & -3.62                                \\ \hline
\end{tabular}}
\caption{\it Best performance described by MSE and NLL  reached for Dropout, DropConnect, RT, and Flipout using both architectures over the test dataset.} \label{table:1}
\end{table}
Several  works  have shown that dropping weights (or neurons) does not bring much performance improvement in convolutional neural networks.  Some authors attribute this   failure   to  the  incorrect  placement  in  the  convolutional blocks~\cite{DBLP:journals/corr/abs-1904-03392},  while others assert that these methods fail in some network architectures~\cite{Gal2015BayesianCN}. Since our aim is searching for a good BNN model useful to analyse the CMB dataset,  hereafter, we will mostly focus  on Dropout for both architectures. 
The analysis show that  large values of Dropout rates are required to decrease the gap between training and validation, and   dropping $10\%$ of the neurons yields the highest performances. Conversely, besides producing better results with respect to AlexNet, VGG also reduces notably the training/validation gap, and only $1\%$ of Dropout rate is required to score the best performance in the model (See Fig.~\ref{fig:4} for AlexNet, and in Fig.~\ref{fig:5} in Appedix~\ref{appendixBhyper}).  We can ascribe this favorable behavior to batch renormalization which acts not only as a regularizer, but also avoids extra normalisation calculations during the forward pass that yield to a quick convergence. On the other hand,  as  discussed in  Sec.~\ref{sec:calibration}, often neural networks are miscalibrated. We then estimate the coverage probabilities corresponding to confidence intervals of   $68\%$, $95.5\%$, and $99.7\%$ (i.e., $1\sigma$, $2\sigma$, and $3\sigma$) in order to verify the accuracy of the uncertainty estimates. The results using AlexNet are shown in table~\ref{table:2}.   We  can  notice  that  firstly as expected the coverage probability is proportional to the Dropout rate, since this variable is associated to the  stochasticity of the model.
\begin{table}[h!]
  \scalebox{0.9}{
\begin{tabular}{|l||l|l|l|l|l|l|l|l|l|l|}
\hline
\multicolumn{11}{|c|}{Coverage probability estimation for Dropout and DropConnect}                                                                                                                                                                                                                                          \\ \hline
\multicolumn{1}{|l||}{\backslashbox{C.I}{Rate}} & \multicolumn{2}{l|}{\textbf{dr=$1\%$}}& \multicolumn{2}{l|}{\textbf{dr=$5\%$}} & \multicolumn{2}{l|}{\textbf{dr=$7\%$}} & \multicolumn{2}{l|}{\textbf{dr=$10\%$}} & \multicolumn{2}{l|}{\textbf{dr=$20\%$}} \\ \cline{2-11} 
\multicolumn{1}{|l||}{}                                                                               & dO.                 & dC.  & dO.                 & dC.                & dO.                & dC.                & dO.                & dC.               & dO.                & dC.               \\ \hline
$68.3\%$                                                                                             & 52.1               &54.9      & 65.0               &66.5             &68.3               & 68.3              & 73.1              &68.4                  &77.1             &76.2              \\ \hline
$95.5\%$                                                                                            & 84.2        &88.2     &93.8           &96.0                   &95.0                   &97.0                   &97.2                   & 98.1                 & 98.4                  & 99.1                 \\ \hline
$99.7\%$                                                                                           &96.3          &96.9     &99.3           &99.8                   &99.4                   &99.6                   & 99.8                  &99.8                  &  99.9                 &99.9                  \\ \hline
\end{tabular}}
\caption{\it Estimation of coverage probabilities corresponding to confidence intervals of  $1\sigma$, $2\sigma$, and $3\sigma$. Dropout rate (dr) becomes a hyper parameter which should be tuned in order to calibrate the network. Dr $\sim0.07$ and dr $\sim0.05$ yield to accurate uncertainties for Dropout and DropConnect, respectively. } \label{table:2}
\end{table}

\begin{table}[h!]
 \scalebox{0.8}{
\begin{tabular}{|l||l|l|l|l|l|l|l|l|}
\hline
\multicolumn{9}{|c|}{Epistemic and Aleatoric uncertainties}                                                                                                                                                                                \\ \hline
\multicolumn{1}{|c||}{\backslashbox{Type of}{Size}} & \multicolumn{2}{c|}{\textbf{$100\%$}}& \multicolumn{2}{c|}{\textbf{$80\%$}} & \multicolumn{2}{c|}{\textbf{$60\%$}} & \multicolumn{2}{c|}{\textbf{$40\%$}}  \\ \cline{2-9} 
\multicolumn{1}{|l||}{Uncertainty}                                                                               & Drop                 & Flip  & Drop                 & Flip                & Drop                & Flip                & Drop                & Flip                          \\ \hline
Epistemic                                                                                            & 0.0024               &0.0011      & 0.0028               &0.0011             &0.0033              & 0.0013             & 0.050              &0.005                              \\ \hline
Aleatoric                                                                                            & 0.090        &0.015     &0.093           &0.014                   &0.094                   &0.019                   &0.14                   & 0.041                                \\ \hline
\end{tabular}}
\caption{\it Aleatoric  and  epistemic  uncertainties  for  a percent  of the total  training  dataset sizes. The aleatoric uncertainty oscillates, while epistemic uncertainty decreases when training dataset gets larger. These results are compatible with those  reported in~\cite{Gal2016Uncertainty}.  There is an anomalous behavior for the $40\%$ of the training subset, which can be explained with the significant overfitting of the model due to the small amount of data.  } \label{table:3}
\end{table}
Interestingly, the Dropout rate that leads to the correct calibration is not  necessarily equal to that one which yields to the best performance. This result supports the fact that calibration of deep neural networks after training becomes the most effective. Additionally, we observe that the Dropout rate used to calibrate  DropConnect models is smaller than the one used in Dropout, suggesting that a stronger stochasticity is involved in  DropConnect networks (since there are more weights than neurons). The behavior in the VGG architecture is substantially different, we observe that is not possible to calibrate the network during training. Tuning the hyper parameters (Dropout rate or posterior regularization) is not enough to tune the uncertainties and thus calibrate the confidence intervals. This is due to the batch (re)normalization layers in the VGG architecture, the normalization applied at each layer indeed standardizes the activation (mean close to 0 and standard deviation close to 1), reducing or even nullifying the effect of the hyper parameters tuning on the epistemic uncertainties.
Re-calibration after training must be applied in networks with batch (re)normalization layers. Therefore, an important result obtained so far is that calibrating networks during training is sometimes not enough, this necessarily  depends on the architecture of the network,  especially if the former contains transformation techniques on the weights like  batch (re)normalization. We  further  computed the  NLL   for different trained dataset size (see  Appendix~\ref{appendixBla}), finding a strong   dependence on  the amount of images used for training the network. 
We also computed the aleatoric and epistemic uncertainties for those experiments (see Table~\ref{table:3}).  Results  show  that  reducing the training dataset size appreciable increases the epistemic uncertainty,   while the aleatoric does not, as discussed in Sec.~\ref{sectII}. Thereby,  observing the effect of the training dataset size on uncertainties should reflect the quality of the uncertainty measurement.

\subsection{ Reparameterization Trick and Flipout}
In this section we evaluate the use of Flipout compared to RT on both  architectures. The performance of both methods are shown in Table~\ref{table:1}. We have found that Flipout outperforms all other methods  regardless the network architecture and also it has achieved significant speedups during the training process. As mentioned above, VGG tends to produce  more miscalibrated networks. This effect was also observed using either Flipout or RT, while for AlexNet we have found out that calibration can be achieved by regularizing the scale parameter  of the approximate posterior of the weights and biases. If the initially trained network is overestimating the error, we want to add a regularization reducing the variance of the approximate posterior. In the case in which the scale of the approximate posterior is parametrized with a softplus function, we can use a SUM regularizer on the parameters of the scale (before the softplus) to reduce the variance and an L2 regularizer to increase it (since the parameters of the scales are negative for a prior with scale around 1).  We also changed  the prior scale, but we have found this approach not effective. The amount of regularization on the parameters of the posterior scale thus play the role the hyper parameter required to calibrate the network.
\begin{table}[h!]
  \scalebox{0.85}{
\begin{tabular}{|l||l|l|l|l|l|l|l|l|l|l|}
\hline
\multicolumn{11}{|c|}{Coverage probability estimation for Flipout and RT.}                                                                                                                                                                                                                                          \\ \hline
\multicolumn{1}{|l||}{\backslashbox{C.I.}{Reg}} & \multicolumn{2}{l|}{\textbf{Non-Reg}}& \multicolumn{2}{l|}{\textbf{Reg=$1e^{-7}$}} & \multicolumn{2}{l|}{\textbf{Reg=$6e^{-6}$}} & \multicolumn{2}{l|}{\textbf{Reg=$1e^{-5}$}} & \multicolumn{2}{l|}{\textbf{Reg=$1e^{-4}$}} \\ \cline{2-11} 
\multicolumn{1}{|l||}{}                                                                               & Flip.                 & RT  & Flip.                 & RT                & Flip.                & RT                & Flip.                & RT               & Flip.                & RT               \\ \hline
$68.3\%$                                                                                             & 63.3               &66.3      & 65.1               &66.5             &68.2               & 67.6              & 70.0              &68.1                  &89.6             &70.0              \\ \hline
$95.5\%$                                                                                            & 91.9        &94.6     &92.9           &94.8                   &95.2                   &95.2                   &95.9                   & 95.5               & 99.4                  & 95.9                 \\ \hline
$99.7\%$                                                                                           &98.8          &99.4     &98.9           &99.4                   &99.4                   &99.6                   & 99.7                  &99.6                 &  100                 &99.6                  \\ \hline
\end{tabular}}
\caption{\it  Estimation of coverage probabilities corresponding to the confidence intervals of  $1\sigma$, $2\sigma$, and $3\sigma$. The regularizer (Reg) becomes an hyper parameter which should be tuned in order to calibrate the network. Reg = $6 e^{-6}$ and Reg= $1e^{-5}$ yield to roughly accurate uncertainties for Flipout and RT respectively. The bias used here is 0.001. } \label{table:2a}
\end{table}

\begin{figure*}[!ht]
  \begin{center}
     \scalebox{0.95}{
\subfloat[$\omega_b$-prediction\label{fig7:preda}]{%
  \includegraphics[width=0.35\textwidth]{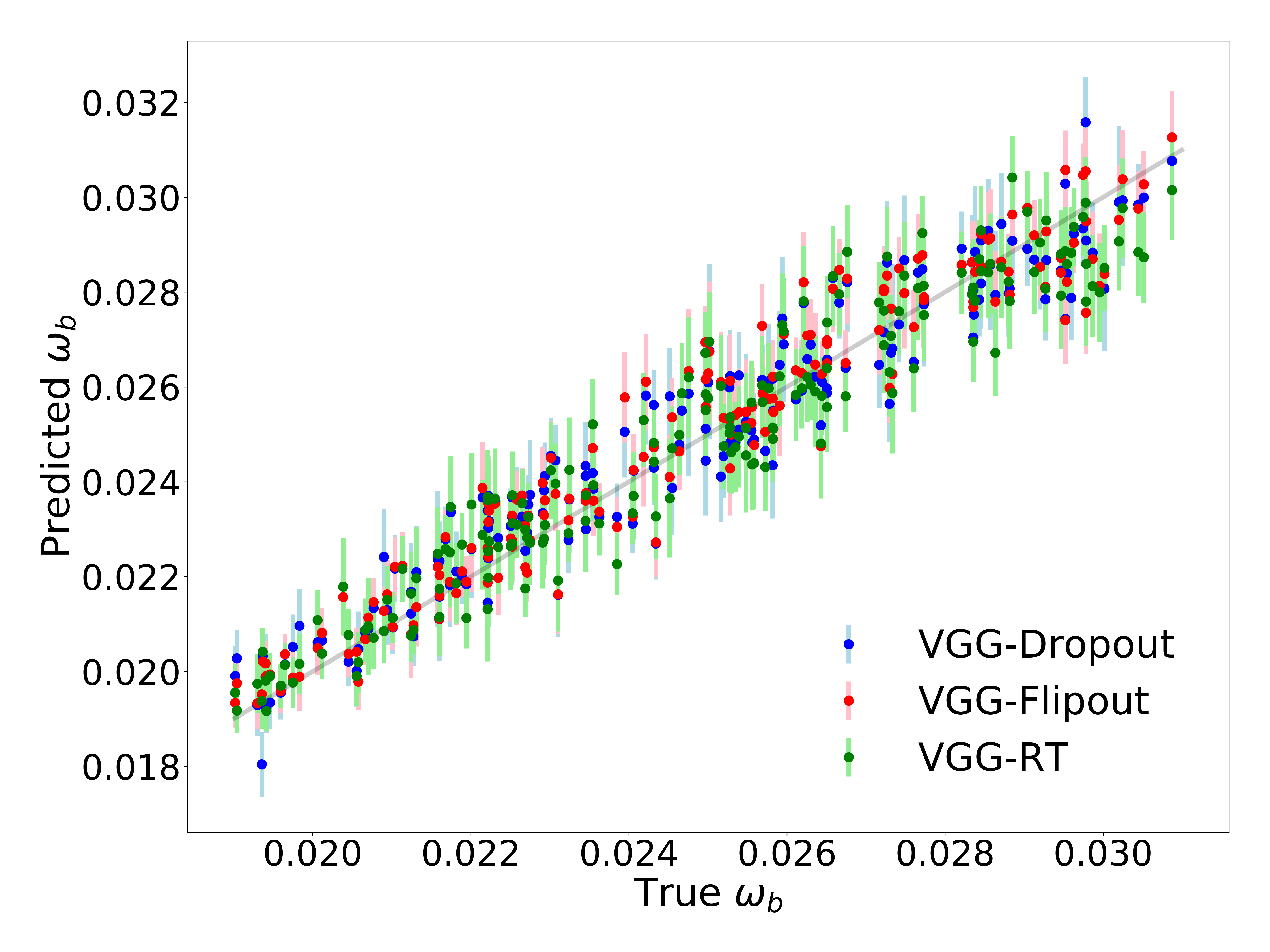}%
}
\subfloat[$A_s$-prediction\label{fig7:predb}]{  \includegraphics[width=0.35\textwidth]{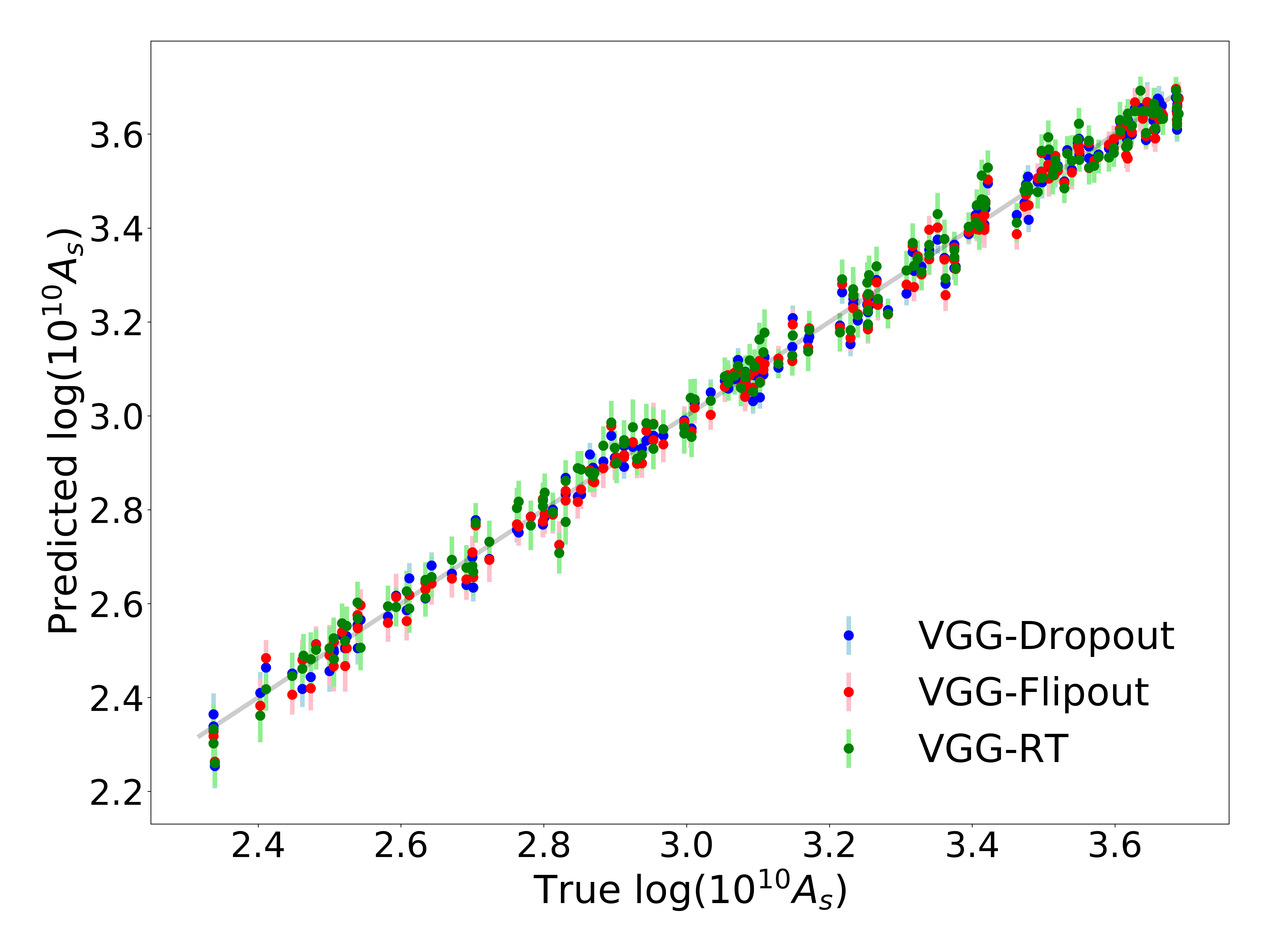}%
}
\subfloat[$\omega_{cdm}$-prediction\label{fig7:predc}]{%
  \includegraphics[width=0.35\textwidth]{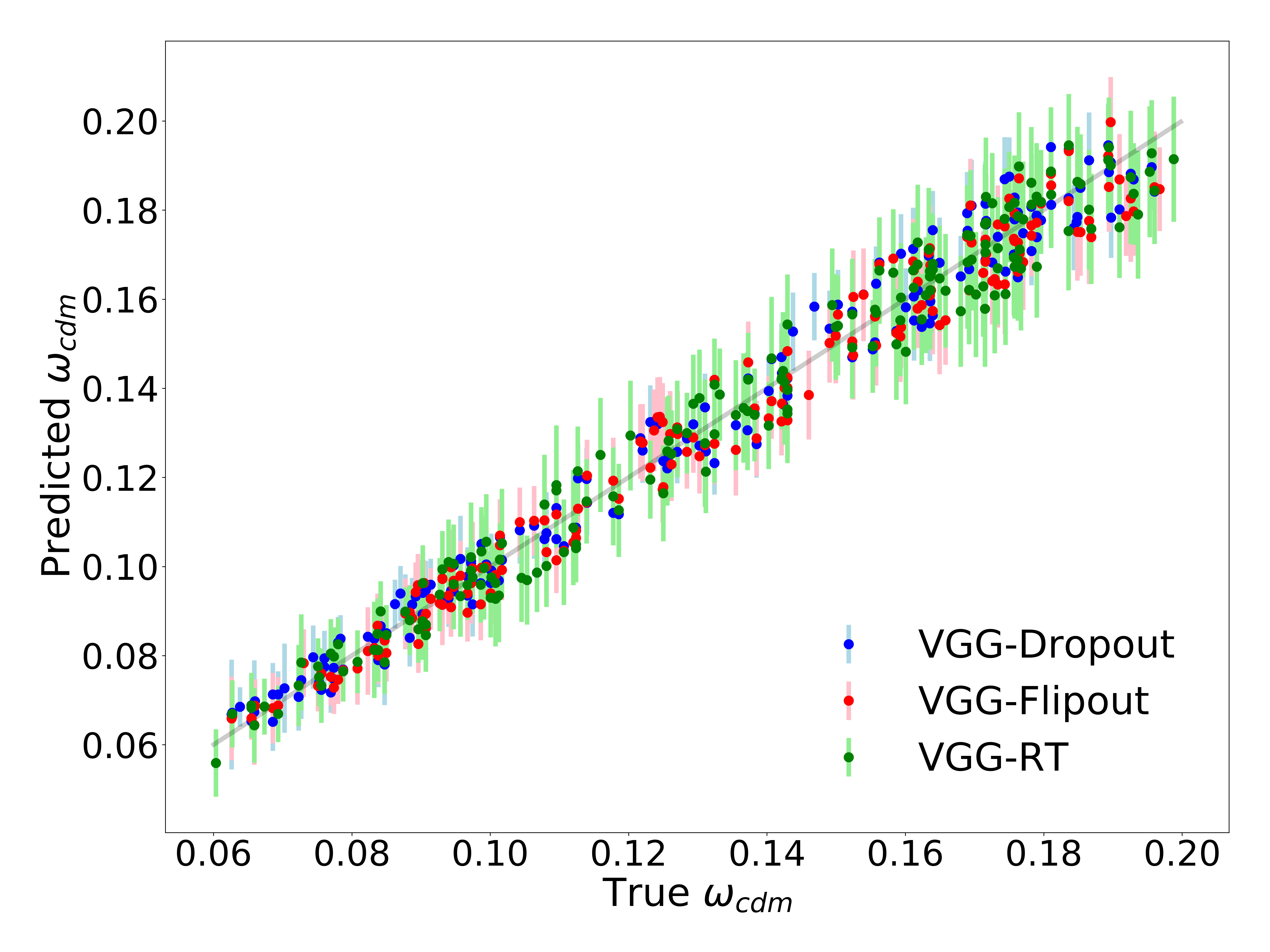}%
}}
\\
\scalebox{0.95}{
\subfloat[$\omega_b$-prediction\label{fig7:preda1}]{%
  \includegraphics[width=0.35\textwidth]{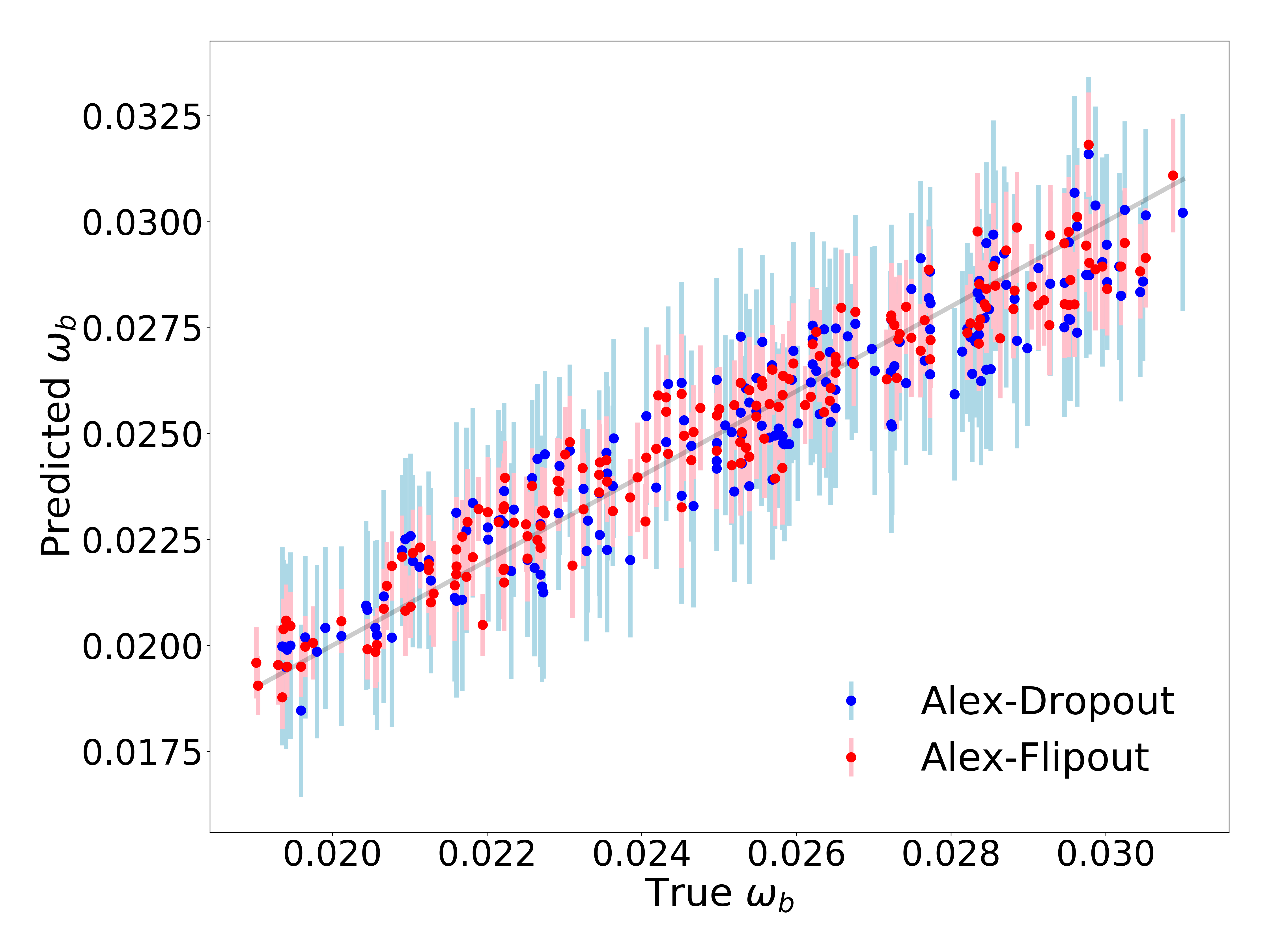}%
}
\subfloat[$A_s$-prediction\label{fig7:predb2}]{%
  \includegraphics[width=0.35\textwidth]{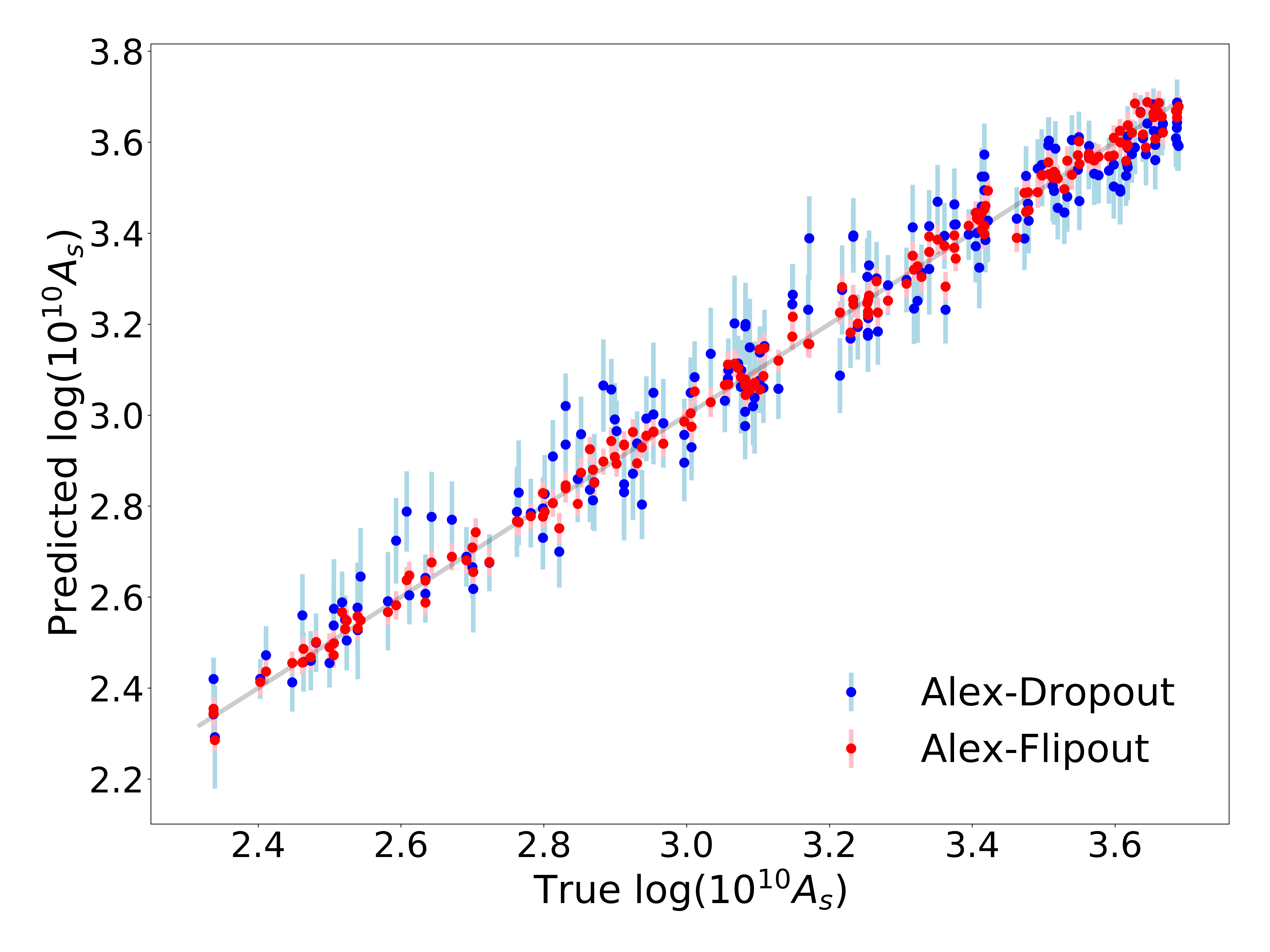}%
}
\subfloat[$\omega_{cdm}$-prediction\label{fig7:predc3}]{%
  \includegraphics[width=0.35\textwidth]{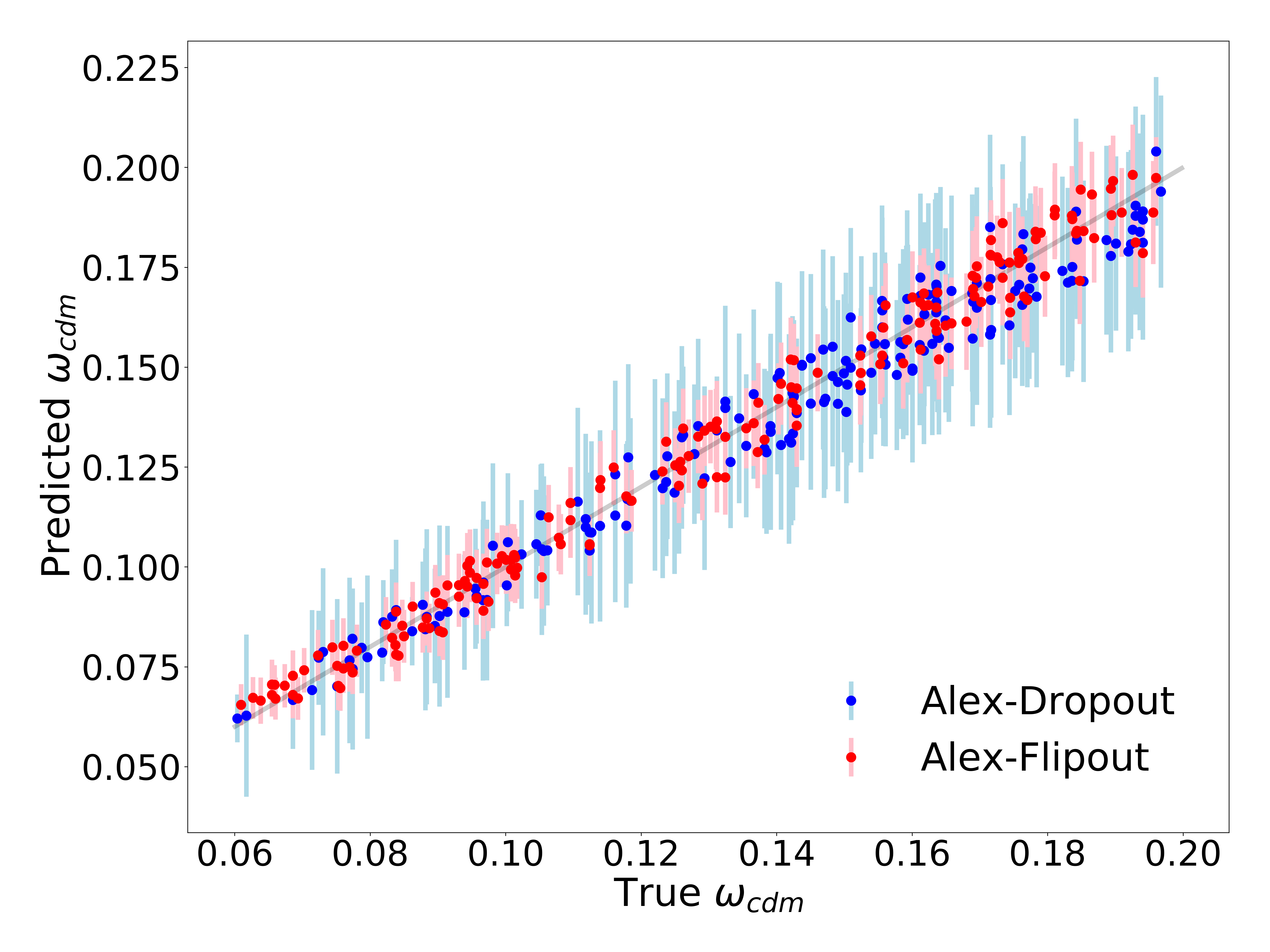}%
}}
\caption{\it  Predictions of the BNN on the test dataset. Top and bottom panels show the results using  VGG and AlexNet, respectively. Each panel shows the parameters used in the dataset and the values predicted from the CMB maps by the neural network. Dots mark the mean of the predictions for 2,500 samples in the test dataset, while the error bars represent their calibrated standard deviations. The colors display different BNNs along with the used architecture.}
\label{fig7:preds}
\end{center}
\end{figure*} 

In fact, regularizing the  posterior allows to reduce the width of the posterior distribution, producing more  accurate  confidence estimates. A visualization of this effect can be seen in Appendix~\ref{appendixBla}.  Table~\ref{table:2a} reports the  coverage probabilities corresponding to confidence intervals of  $1\sigma$, $2\sigma$, and $3\sigma$ for different values of the regularization. We have found out that without any regularizer, the estimation of the error is permissive and  enhancing  this hyper parameter increases the coverage probability estimation until arriving at values very close to their corresponding confidence intervals.  The values reached to calibrate the network are  $\sim 6 e^{-6}$ and $\sim 1e^{-5}$  for Flipout and RT, respectively. Fig.~\ref{fig:1a} in Appendix~\ref{appendixBhyper} displays the performance of the network for the models used in Table~\ref{table:2a} only for Flipout.
Despite the fact that BNNs incorporate some degree of  regularization,  the gap between training/validation still remains for AlexNet architecture, while for VGG it becomes small. Additionally, we can estimate both the epistemic and aleatoric uncertainties from the calibrated Flipout network. The results are shown in  Table~\ref{table:3}. As before, epistemic uncertainty increases with the size of the training dataset. However, we do observe that the epistemic uncertainties becomes smaller for Flipout compared to Dropout, implying that Flipout indeed achieves the largest variance reduction. The performance of the network using different training dataset sizes can be seen in Appendix~\ref{appendixBla}. Although the use of batch (re)normalization  produces reduction on the training/validation gap, it  also leads to large fluctuations due to the fact that it is constantly readjusting the layers to new distributions.   
\subsection{\label{sectapprox} Approximated the Posterior Distribution  of the Cosmological Parameters }

\begin{figure}[h!]
\begin{center}
\includegraphics[width=0.5\textwidth]{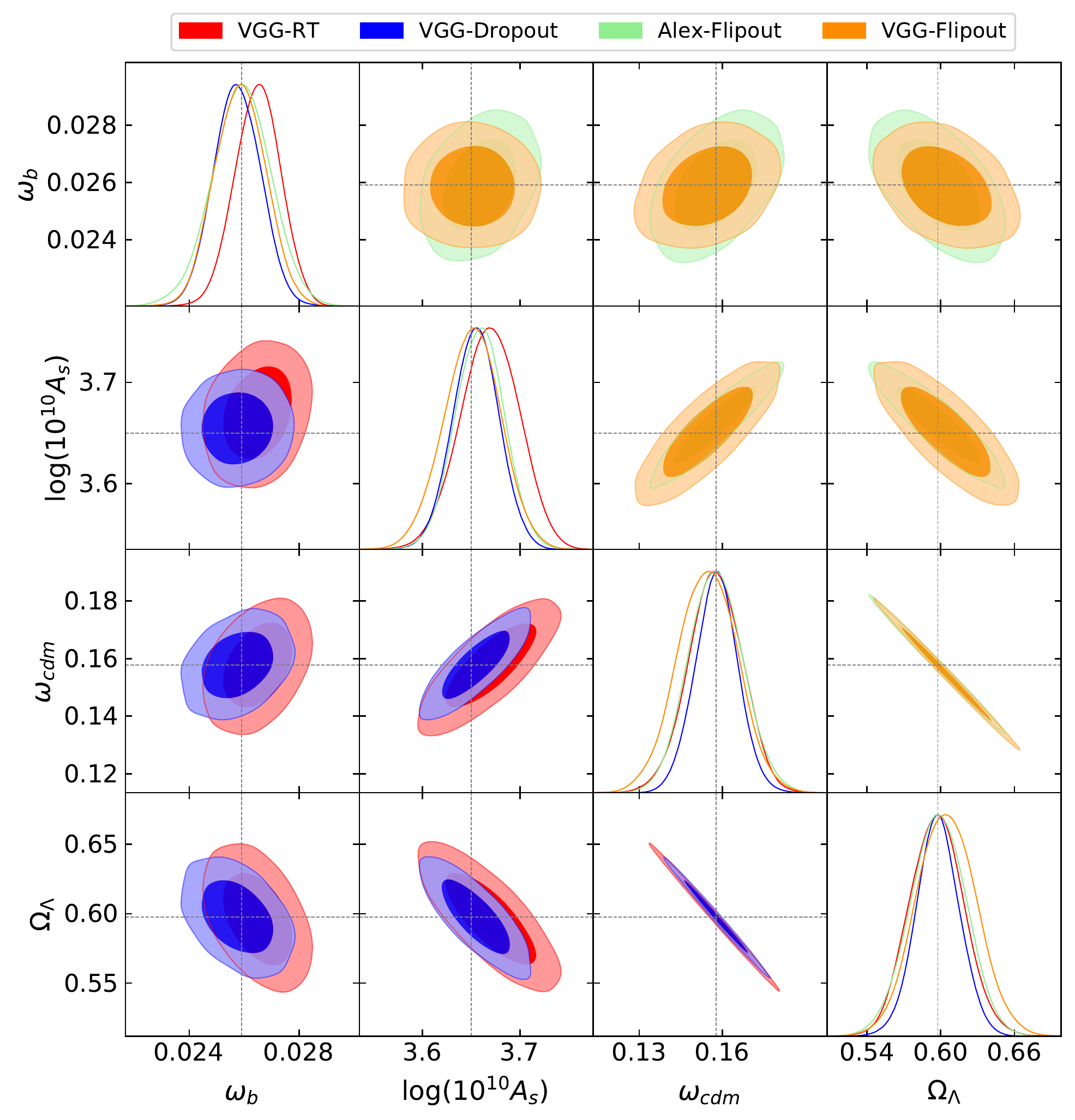}
\end{center}
\caption{\it  Minimal base-$\Lambda$CDM $68\%$ and $95\%$ parameter constraint contours from one example of our synthetic CMB dataset using the best among four methods. The diagonal plots are the marginalized parameter constraints, the dashed lines stand for the predicted values and the black solid line corresponds to the true values $\omega_b=0.02590$, $\log(10^{10}A_s)=3.65653$ and  $\omega_{cdm}=0.15773$. } \label{fig:triang2}
\end{figure}
So far we have analyzed the performance of different BNNs for each architecture. We have found out that Flipout and Dropout are methods which work really well to carry out  parameter inference using our CMB dataset. After calibrating the network with the approach introduced  in Sec.~\ref{sec:calibration}, we can visualize the performance of the above methods  in  terms  of  precision  of  their  cosmological  parameters predictions. Fig.~\ref{fig7:preds} shows the correlation between the true parameters and the predicted parameters with the respective confidence intervals, on the test CMB maps.
As mentioned in the previous section, implementing Dropout in the AlexNet architecture leads to large MSE values and larger uncertainties, while  Flipout keeps excellent performance regardless the architecture, and its uncertainties are  notably reduced.  Finally, we constraint the dark-energy density written as  $\Omega_{\Lambda}\approx 1-\Omega_b-\Omega_{cdm}$, being $\omega_i\equiv\Omega_{i}h^2$ with $h=0.6781$~\cite{Aghanim:2018eyx}, just to examine the posterior distribution for the derived parameters. The triangle plots of Figs.~\ref{fig:triang2} and~\ref{fig:triang1} show our main results for one example randomly picked from the CMB test dataset. We use the Python GetDist package for creating the triangular plot~\cite{lewis2019getdist}. Fig.~\ref{fig:triang2} displays the four most accurate BNNs  which yield credible cosmological parameter contours. The target values for the selected example are  $\omega_b=0.02590$, $A_s=3.65653$,  $\omega_{cdm}=0.15773$, the derived parameter $\Omega_\Lambda=0.59761$, and in   Table~\ref{table:parameters} gives its marginalized parameter constraints from the  CMB maps. In Appendix~\ref{appendixA} we show  the results for all the BNNs  introduced in this paper.

\begin{table}[h!]
  \scalebox{0.8}{
\begin{tabular}{|l||l|l|l|l|l|}
\hline
\multicolumn{5}{|c|}{Marginalized parameter constraints}                                                                                                                                                                                                                                          \\ \hline
\multicolumn{1}{|l||}{\backslashbox{$\Lambda$CDM}{BNN}} & \multicolumn{1}{l|}{\textbf{Flipout-Alex}}& \multicolumn{1}{l|}{\textbf{Flipout-VGG}} & \multicolumn{1}{l|}{\textbf{RT-VGG}} & \multicolumn{1}{l|}{\textbf{Drop-VGG}} \\ \cline{2-5} 
\hline\rule{0pt}{10pt}
$ \omega_{b}$                                           & $0.0259^{+0.0021}_{-0.0021}$ &  $0.0259^{+0.0017}_{-0.0018}$ & $0.0265^{+0.0016}_{-0.0016}$ &  $0.0258^{+0.0017}_{-0.0016}$                         \\ \hline
\rule{0pt}{10pt}$\ln(10^{10}A_s)$                                                   &$3.659^{+0.048}_{-0.049}    $    &$3.650^{+0.057}_{-0.056}   $ & $3.669^{+0.059}_{-0.063}   $   &               $3.655^{+0.046}_{-0.045}   $                          \\ \hline \rule{0pt}{10pt}
$\omega_{cdm}$                                           & $0.158^{+0.019}_{-0.019}   $   &$0.155^{+0.021}_{-0.021}   $ & $0.157^{+0.019}_{-0.019}   $   &          $0.158^{+0.014}_{-0.015}   $                            \\ \hline\rule{0pt}{10pt}
$\Omega_{\Lambda}$                                        &  $0.598^{+0.045}_{-0.043}   $  &  $0.604^{+0.046}_{-0.047}   $ &  $0.598^{+0.043}_{-0.042}   $   &                             $0.597^{+0.034}_{-0.032}   $          \\ \hline
\end{tabular}}

\caption{\it Parameters $95\%$ intervals for the minimal base-$\Lambda$CDM model from our synthetic CMB dataset using Flipout, RT, and Dropout with the AlexNet and VGG architectures. } \label{table:parameters}
\end{table}

\subsection{\label{secpol} Parameter Estimation from Combination of Temperature and Polarization Maps}

\begin{figure*}[!ht]
  \begin{center}
     \scalebox{0.95}{
\subfloat[$\omega_b$-prediction\label{fig7:predapol}]{%
  \includegraphics[width=0.35\textwidth]{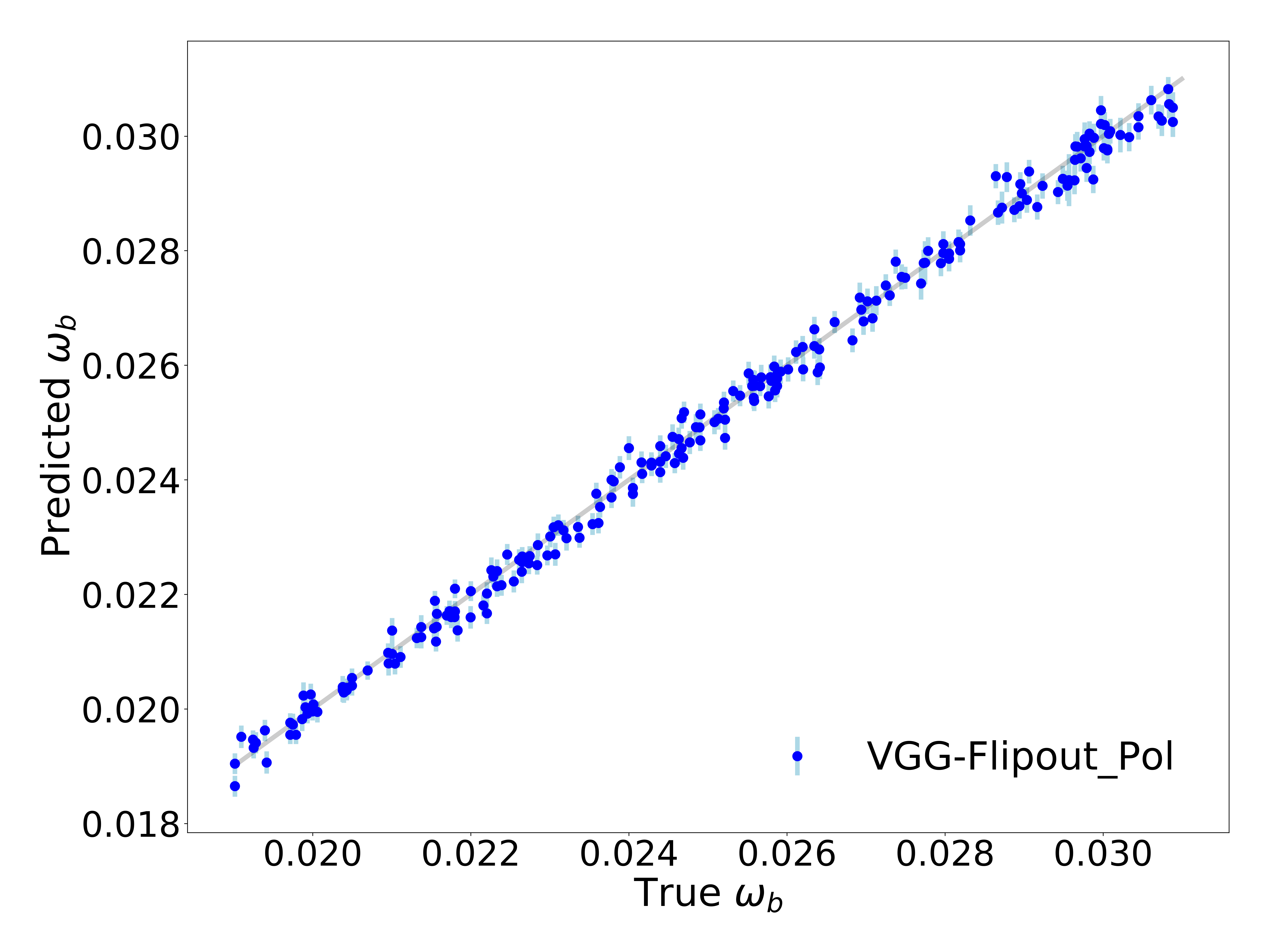}%
}
\subfloat[$A_s$-prediction\label{fig7:predbpol}]{  \includegraphics[width=0.35\textwidth]{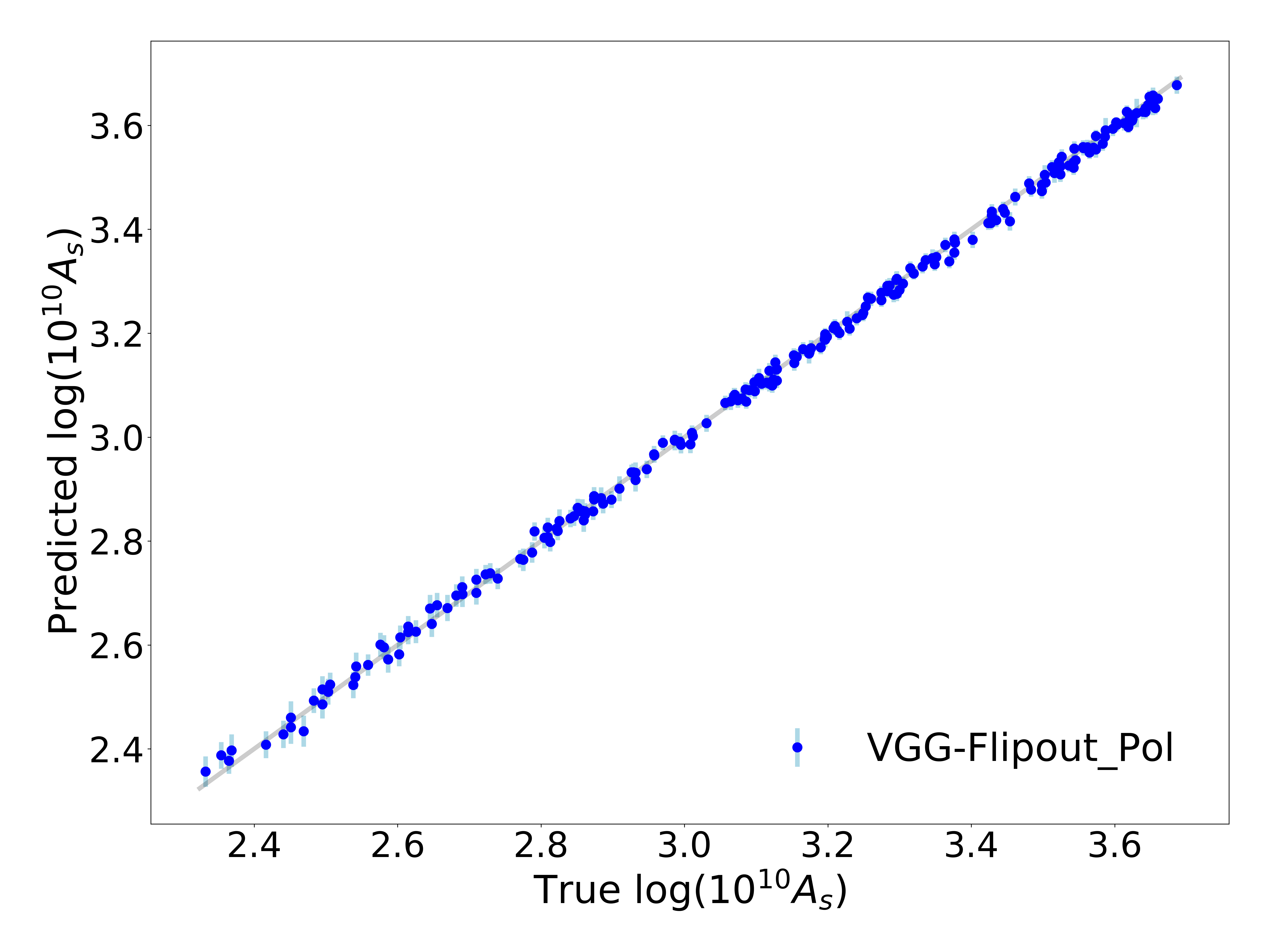}%
}
\subfloat[$\omega_{cdm}$-prediction\label{fig7:predcpol}]{%
  \includegraphics[width=0.35\textwidth]{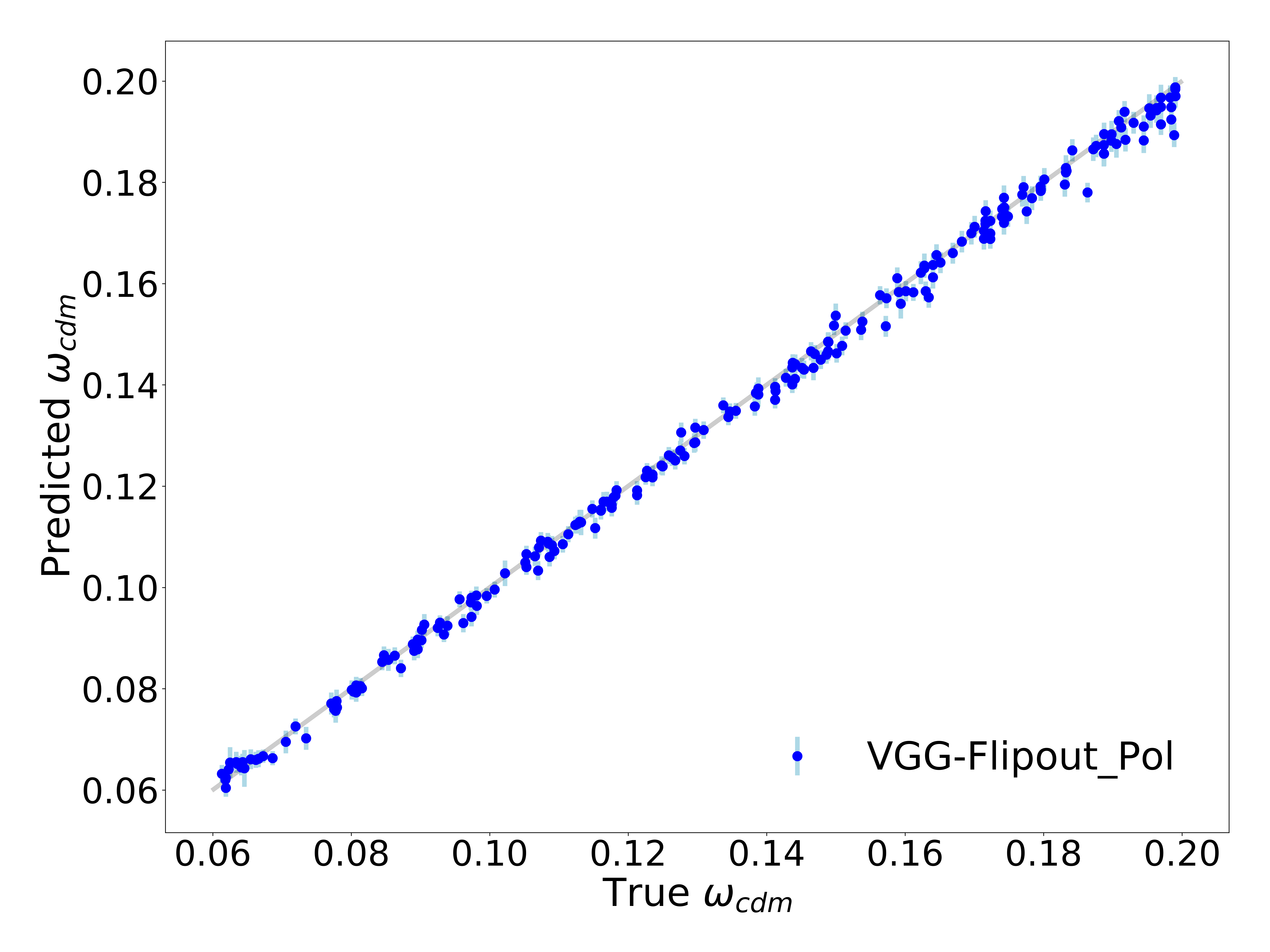}%
}}

\caption{\it  Predictions of the BNN on the test dataset. The plots  show the parameters used in the dataset and the values predicted from the CMB temperature plus polarization maps by the neural network. Dots mark the mean of the predictions for 2,500 samples in the test dataset, while the error bars represent their calibrated standard deviations. Here we use the Flipout method.}
\label{fig7:predspol}
\end{center}
\end{figure*} 

\begin{figure}[h!]
\begin{center}
\includegraphics[width=0.47\textwidth]{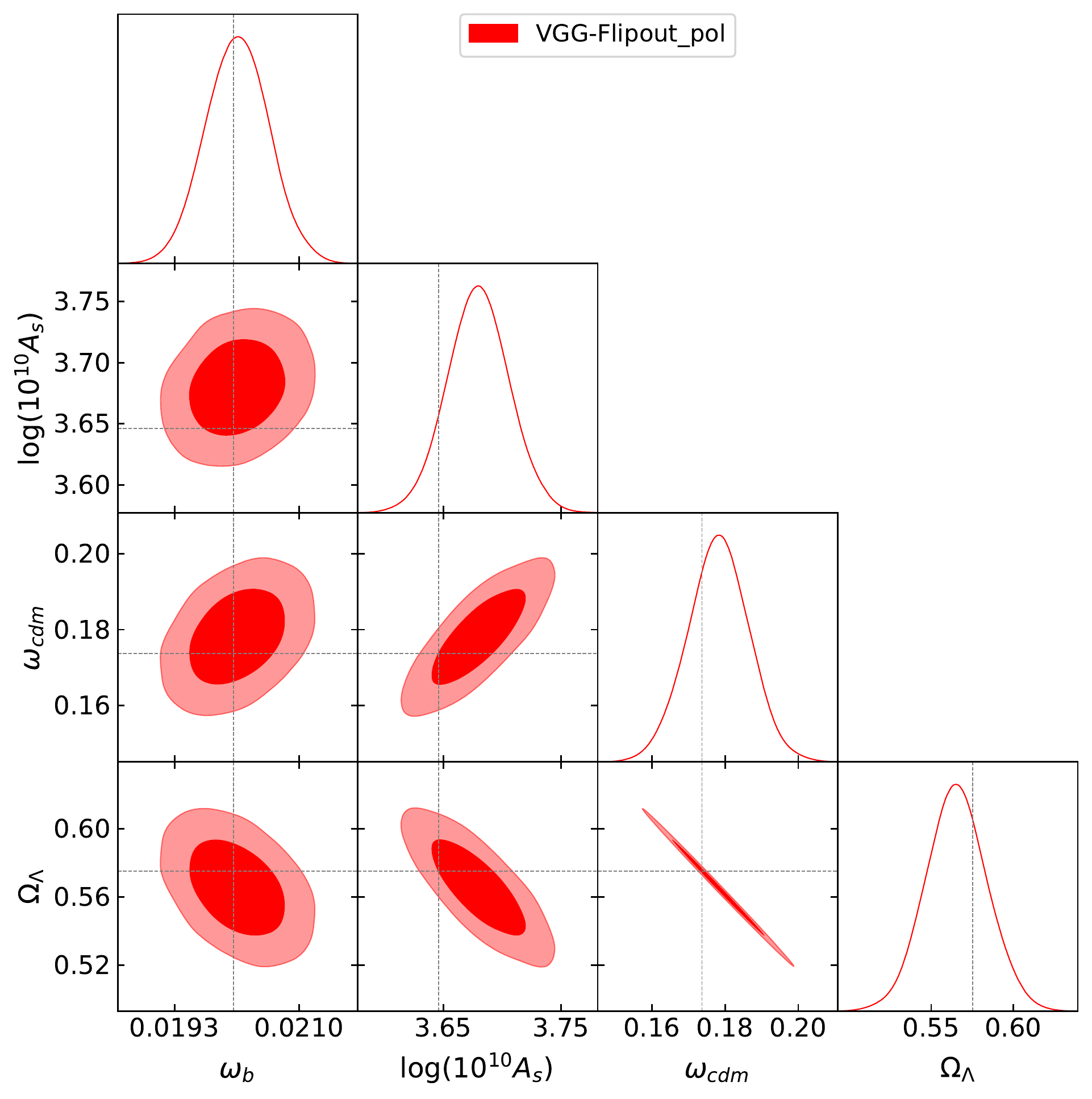}
\end{center}
\caption{\it  Minimal base-$\Lambda$CDM $68\%$ and $95\%$ parameter constraint contours from one example of our synthetic CMB dataset with temperature and polarization  using Flipout. The diagonal plots are the marginalized parameter constraints, the dashed lines stand for the predicted values and the black solid line corresponds to the true values. } \label{fig:triangpol}
\end{figure}
So far, we have obtained the parameter predictions only from maps of CMB temperature. Adding complementary information to the dataset such as a polarization,  will provide better  constraints on the cosmological model  and  also help to break some partial parameter degeneracies. In  this subsection we present predictions of the parameters from the combinations of temperature and polarization which are significantly more precise that those determined using temperature alone.\\
Cosmological experiments like Planck are designed  to  measure the Stokes  parameters $T$, $Q$, $U$ useful for  analyzing the CMB radiation in terms of its temperature  and  polarization. This CMB radiation field can be expressed as a rank-2 tensor $\mathcal{I}_{ij}$, where $T=(\mathcal{I}_{11}+\mathcal{I}_{22})/4$ corresponds to the temperature anisotropy studied in the previous sections, while $Q=(\mathcal{I}_{11}-\mathcal{I}_{22})/4$ and $U=\mathcal{I}_{12}/2$ describe the linear polarization~\cite{Zaldarriaga_1998}. 
While $T$ is a scalar invariant under rotations, $Q$ and $U$ depend on the reference frame determined  by the direction of observation  ${\bf \hat{n}}$ and  two axes $(\hat{{\bf e}}_1,\hat{{\bf e}}_2)$ perpendicular to ${\bf \hat{n}}$.
If $\hat{{\bf e}}_1$ and $\hat{{\bf e}}_2$ are rotated by an angle $\psi$, $Q$ and $U$ transform as~\cite{zaldarriaga2003polarization}
\begin{equation}
    (Q\pm iU)^\prime({\bf\hat{n} })=e^{\mp 2i\psi}(Q\pm iU)({\bf\hat{n} }),
\end{equation}
where  the prime denotes the quantities in the transformed coordinate system.  For these objects, one can construct two real  quantities usually called $E$ and $B$  which are invariant under rotations, but they behave differently under parity ($E$ remains unchanged, while $B$ changes sign)~\cite{KAPLAN2003917}. Therefore, adding   polarization  and  assuming that  CMB fluctuations are Gaussian, the statistical properties of the CMB in the sky  are fully encoded in  four power spectra $C^{TT}$, $C^{EE}$, $C^{BB}$, and $C^{TE}$. \\
In order to estimate the cosmological parameters directly from the CMB maps, we run again the script used in Sec.~\ref{sectV} but this time, generating the four power spectra with the CLASS code. Those spectra are given in input to healpy which produces three maps associated for: $T$, $Q$, and  $U$. Finally, these three maps are stacked in order to create images of size $256\times256\times3$, analogous to the RGB images, where each channel corresponds to a map measured by the cosmological experiment. We choose the Flipout method for estimating the cosmological parameters from temperature and polarization maps due to the notable performances found previously. We use the VGG architecture since yield the best results for the temperature map alone, and we calibrate the network after training with the usual method of Sec.~\ref{sec:calibration}.
\begin{table}[h!]
  \scalebox{0.8}{
\begin{tabular}{|l||l|l|l|l|l|}
\hline
\multicolumn{5}{|c|}{Metrics for the Network with Polarization}  \\ \hline
\multicolumn{1}{|l||}{\backslashbox{Metrics}{Size}} & \multicolumn{1}{l|}{$100\%$}& \multicolumn{1}{l|}{$80\%$} & \multicolumn{1}{l|}{$60\%$} & \multicolumn{1}{l|}{$40\%$} \\ \cline{2-5} 
\hline\rule{0pt}{10pt}
Epistemic           & $9.50e^{-5}$ &  $10.01e^{-5}$ & $9.51e^{-5}$ &  $10.40e^{-5}$          \\ \hline
\rule{0pt}{10pt}Aleatoric   &$1.2e^{-3}$  &  $1.3e^{-3}$ & $1.2e^{-3}$   & $1.9e^{-3} $   \\ \hline \rule{0pt}{10pt}
NLL         & $-5.932 $   &$-5.911  $ & $-5.897   $   &          $-5.380  $                            \\ \hline\rule{0pt}{10pt}
MSE         &  $0.0034  $  &  $0.0034  $ &  $0.0036   $   &         $0.0060  $          \\ \hline\rule{0pt}{10pt}
C.I-$1\sigma$ &$68.6\%$ &$67.9\%$&$67.1\%$&$71.5\%$ \\ \hline\rule{0pt}{10pt}
C.I-$2\sigma$ &$94.7\%$ &$94.7\%$&$93.1\%$&$94.6\%$ \\ \hline\rule{0pt}{10pt}
C.I-$3\sigma$ &$98.8\%$ &$98.7\%$&$98.9\%$&$99.2\%$ \\ \hline
\end{tabular}}
\caption{\it  Assessment  of Flipout model using  synthetic CMB temperature and polarization maps. Below  is reported the Confidence Interval results for $68\%$, $95.5\%$ and $99.7\%$. } \label{table:pol1}
\end{table}
We made several experiment varying the size of the training set, the results are shown in Table~\ref{table:pol1}. Despite of the results obtained in Table~\ref{table:3}, we can observe how epistemic and  aleatoric remains approximately constant. In this case, thanks to the additional information provided by the polarization channel, even the $40\%$ percent of the data seems to be enough for training the network properly reaching convergence of the uncertainties. With respect to the temperature map alone, the values for both uncertainties decrease one order of magnitude and the gap between them is also reduced. Furthermore, the MSE decreases almost one order of magnitude with respect to the values found in Subsec.~\ref{sectapprox}. The predictions for each cosmological parameter are displayed in Fig.~\ref{fig7:predspol}. Evidently, we can observe the improvements provided by polarization, in particular the reduction of uncertainty intervals for all  parameters. In order to show the parameter intervals  and contours  from  the  combined CMB and polarization map, we choose randomly an example from the test set  with true values $\omega_b=0.0201$, $\log(10^{10}A_s)=3.6450$ and  $\omega_{cdm}=0.1736$. The two-dimensional posterior distribution of the cosmological parameters are shown in Fig.~\ref{fig:triangpol} and the parameter $95\%$ intervals  are given by
\begin{eqnarray}
&&\omega_b= 0.02015^{+0.00086}_{-0.00086},\quad \log(10^{10}A_s)=3.680^{+0.052}_{-0.052}\nonumber\\ &&\omega_{cdm}=0.178^{+0.016}_{-0.016},\quad \Omega_\Lambda= 0.565^{+0.037}_{-0.037}
\end{eqnarray}
Comparing Figures~\ref{fig:triangpol} and \ref{fig:triang2} we can observe that adding polarization data provides considerably tighter constraints on all  parameters than can be obtained from only temperature data.  Presence of polarization information can  break degeneracies among  parameters and stringent its constraints which is   consistent with our preliminary power spectrum analysis in Sec.~\ref{sec:MCMC}. The parameter degeneracy that determines the  incapacity to distinguish certain  parameter combinations, emerges not only  from physical effects such as a  geometrical degeneracy~\cite{10.1046/j.1365-8711.1999.02274.x}, but also from low numerical precision in the methods used for computing the cosmological observables~\cite{Howlett_2012}. In this case, the distinctive effects that $\omega_b$ produces in temperature and polarization spectra allows the network to recognize the impact of this parameter in the maps.\\

\subsection{\label{secpolcal} Calibrating Bayesian Networks via Gradient Descent}
Temperature Scaling (TS) is one of the most appealing method  used in literature for calibration~\cite{2019arXiv190511659L}. It consists in finding the value of a scalar parameter on  validation set such that minimize NLL. We argue that this method can only reduce aleatoric uncertainties, while it cannot reduce epistemic uncertainties in BNNs driven by the  means of the prediction distributions.
To this aim, we propose a technique that re-adjust the weights in the last layer of the network which are associated to both means and covariance of the output distribution. We call Last Layer  the method that affects  all last layer, while we call Last Layer Loc the method affecting only the weights associated with the prediction means. The NLL function Eq.~\eqref{eq:16} can be transformed in two ways: $\Sigma\rightarrow s\Sigma$  being $s\in\mathbb{R}^+$ a scalar parameter,(which is the TS method)  and $\Sigma\rightarrow L \Sigma L^\top$  being $L$ a lower triangle matrix. In order to verify the reliability of these methods, we load the weights computed from the model reported in Subsec.~\ref{secpol} and we optimize the NLL for 100 epochs after convergence. The obtained results are reported in Table~\ref{table:polcal}. We can observe that for the last layers methods (LL and LL-loc) the NLL and MSE is reduced significantly with respect to the values found in Table~\ref{table:pol1} (100\% column), and the aleatoric uncertainty is also notably reduced for all methods.

\begin{table}[h!]
  \scalebox{0.8}{
\begin{tabular}{|l||l|l|l|l|l|l|l|}
\hline
\multicolumn{7}{|c|}{Metrics for our calibrated methods}\\ \hline
\multicolumn{1}{|l||}{\backslashbox{Metrics}{Methods}} & \multicolumn{1}{l|}{$LL_S$}& \multicolumn{1}{l|}{$LL_T$} & \multicolumn{1}{l|}{$LL-loc_S$} & \multicolumn{1}{l|}{$LL-loc_T$}&
\multicolumn{1}{l|}{$TS_S$} & \multicolumn{1}{l|}{$TS_T$}\\
\cline{2-7}
\hline\rule{0pt}{10pt}
Epistemic   & $2.4e^{-4}$ &  $2.4e^{-4}$ & $2.4e^{-4}$ &  $2.4e^{-4}$ &$2.4e^{-4}$ & $  2.4e^{-4}$                      \\ \hline
\rule{0pt}{10pt}Aleatoric          &$ 9.6e^{-4} $    &$ 9.7e^{-4} $ & $ 9.6e^{-4}  $   &               $ 9.2e^{-4} $            &$ 12.3e^{-4}  $ &    $ 11.1e^{-4}  $          \\ \hline \rule{0pt}{10pt}
NLL                                           & $-6.71 $   &$-6.67  $ & $-6.31   $   &          $-6.50  $              &$-6.06$ &$-6.13$             \\ \hline\rule{0pt}{10pt}
MSE   &  $0.0026  $  &  $0.0026  $ &  $0.0027   $   &  $0.0026  $   &$0.0033$ &$0.0032$      \\ \hline\rule{0pt}{10pt}
C.I-$1\sigma$ &$79.1\%$ &$79.1\%$&$80.0\%$&$79.6\%$&$79.0\%$ &$77.2\%$ \\ \hline\rule{0pt}{10pt}
C.I-$2\sigma$ &$97.8\%$ &$97.8\%$&$98.5\%$&$98.0\%$&$98.3\%$ &$97.7\%$ \\ \hline\rule{0pt}{10pt}
C.I-$3\sigma$ &$99.9\%$ &$99.9\%$&$99.9\%$&$99.9\%$&$99.9\%$ &$99.6\%$ \\ \hline
\end{tabular}}
\caption{\it  Assessment  of Flipout model using  synthetic CMB temperature and polarization maps. In this case, the calibration was achieved using three proposal methods: Temperature Scaling (TS), Last Layer (LL) and Last Layer  only with the mean (LL-loc). In each case, the temperature scaling parameters are either a single scalar (S) or a lower triangular matrix (T). We report also the Confidence Interval results for $1\sigma$, $2\sigma$ and $3\sigma$.} \label{table:polcal}
\end{table}

\begin{figure}[ht!]
\begin{center}
\includegraphics[width=0.47\textwidth]{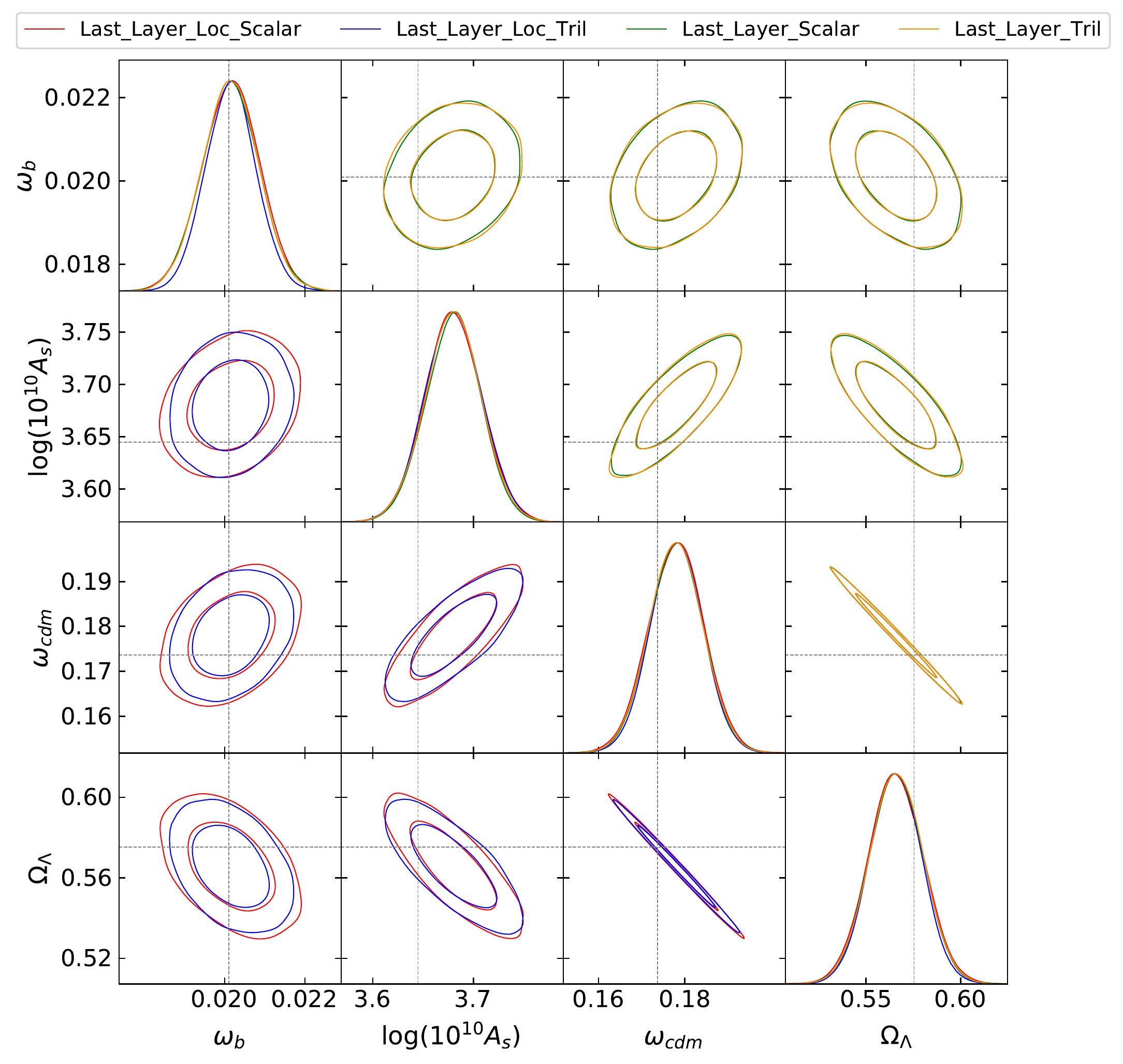}
\end{center}
\caption{\it  Minimal base-$\Lambda$CDM $68\%$ and $95\%$ parameter constraint contours from one example of our synthetic CMB dataset using Flipout. In this case, the calibration was achieved using two proposal methods: Last Layer  and Last Layer  only with the mean(Loc). In each case, the optimizer parameter is either a scalar  or a matrix (Tril).} \label{fig:triangpolcal}
\end{figure}
Minimizing a scalar produce the same orientations that we found with the method used in the previous sections, while the use of lower triangular matrix as a scaling reorient a bit the ellipses during calibration. Unfortunately all methods seem to produce poor estimations for the confidence intervals. However, preliminary results have shown that some techniques combined  with TS leads to  more reliable uncertainty estimates~\cite{hortua2020reliable}. One of these techniques is related to the generalization of the KL-divergence used in VI, Eq.~\ref{eq:4}. Using approximate approaches such BB-$\alpha$~\cite{hernndezlobato2015blackbox}, we found  that  some divergences related to  Hellinger distances  or  power Expectation Propagation (EP) method yield to well-calibrated networks.  Fig.~\ref{fig:triangpolcal}  shows the contours at $68\%$ and $95\%$   using the example shown in the previous subsection. Here we have used the combination of our proposal method with the BB-alpha  where its hyper-parameters were adapted to   work with the power EP method.  For all methods, there is no distinction between using temperature scaling scalar (S) or  a lower triangular matrix (T) finding  its consistently outputs with well-calibrated credible intervals. More detailed analysis about these calibration proposals will be the object of future works~\cite{hortua2020reliable}.

\section{\label{sectVIII}Conclusions}
We have employed Bayesian neural networks as a reliable and accurate tool to estimate the 
posterior distribution of the cosmological parameters directly from simulated CMB maps. BNNs, when properly trained and calibrated, offer the capability to estimate the total uncertainty (aleatoric and epistemic) of their predictions. They are trained by imposing a prior distribution on the parameters of the network (weights) and approximating the posterior distribution using Variational Inference. Different assumptions about the distribution over the weights have been proposed, and four of them were used in this work: Dropout, DropConnect, Flipout, and  Reparameterization Trick. We have compared them by implementing two architectures and several calibration methods.

\vspace{0.5cm} {{\bf Takeaway messages:}}
\begin{enumerate}
\item As a proof-of-principle,  we  have compared  Variational Inference(VI) and   MCMC techniques finding that the latter excels at quantifying uncertainty, while the former  is about 10000 times faster at inference. Using the covariance matrix efficiently estimated from the BNNs samples as initial proposal in MCMC, significantly increases the acceptance rate and gives faster convergence.
 \item Flipout emerged as the most reliable and effective method, achieving best performances across architectures, while DropConnect had the worst performances. Furthermore, we observed that Flipout converges much faster during training and manifests a notable reduction in the credible contours for the parameters.
 \item  Calibrating after training becomes the best option. In fact, we showed that hyper parameter tuning in training is not sufficient in the cases where batch (re)normalization is present in the architecture. Therefore, calibration in these architectures is only possible after training. Using batch (re)normalization is advantageous since it allows us to obtain the best performances and the highest convergence rate during training while focusing on calibration after.
 \item We observed  that tuning the regularization parameter for the scale of the approximate posterior on the weights in Flipout and RT we can produce
   unbiased and reliable uncertainty estimates, similar to  Dropout rate in MCDropout.
 \item Parametric methods for  calibration introduced in Sec.~\ref{sectVII}  are simple but quite successful. Different calibration methods based on gradient descent have also been proposed, to be used alone or in combination with Temperature Scaling. Further investigation on these calibration methods will be the object of future studies.
 \item  In Subsec.~\ref{secpol} we show how polarization can be combined with the temperature in a unique multi-channel input tensor, allowing the network to automatically extract complex information from partial sky coverage maps and significantly reducing the prediction errors (see  Fig.~\ref{fig:triangpol} and Table~\ref{table:pol1}).  The outcomes of this subsection  establish a remarkable result that will be further explored in subsequent studies.
 \end{enumerate}

The research showed in this paper  allows to   extract  relevant features  directly from the raw data such as non-Gaussian signals~\cite{2015JCAP...09..064N} or foreground emissions \cite{NrgaardNielsen2010ForegroundRF,doi:10.1002/asna.201813428}. In future work, we plan to carry out this comparison improving the architecture and using other cosmological datasets such as large-scale matter distribution or $21$cm maps.  The data generator script and MCMC chains given here are  available at  \footnote{\url{ https://github.com/JavierOrjuela/BayesianNeuralNets_CMB}}.  The library  Argo used for obtaining the results reported in this paper can be found at~\footnote{\url{https://github.com/rist-ro/argo}}.

\begin{acknowledgments}
 H.J.~Hort\'ua, R.~Volpi, and L.~Malag\`o are supported by the DeepRiemann project, co-funded by the European Regional Development Fund and the Romanian Government through the Competitiveness Operational Programme 2014-2020, Action 1.1.4, project ID P\_37\_714, contract no. 136/27.09.2016. 
D.~Marinelli acknowledges the RIST institute where he was employed when this projected started and the initial support from the DeepRiemann project.

\end{acknowledgments}

\nocite{*}

\bibliography{paper}

\appendix
\section{Evaluation of  coverage probabilities through  binned samples}
\label{appendixA1}
As mentioned in Sec.~\ref{sectV}, if the distribution that describes the samples drawn from the posterior  is Gaussian, we can compute the coverage probabilities from the ellipsoidal confidence. 
\begin{figure}[h!]
\begin{center}
\includegraphics[width=0.45\textwidth]{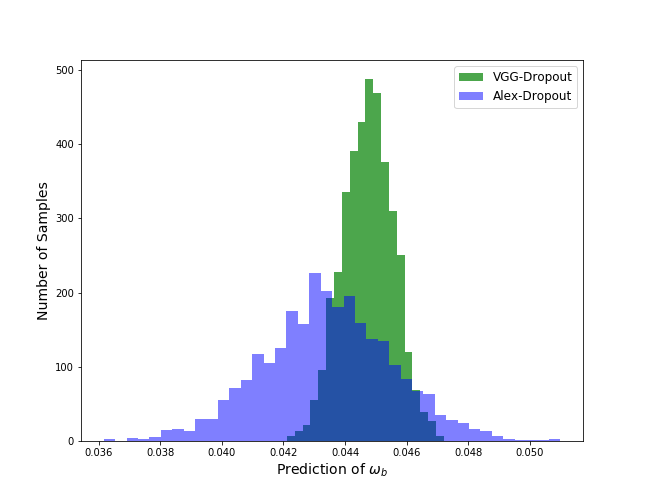}
\end{center}
\caption{\it  Histogram generated from binned samples drawn from the posterior of the parameter $\omega_b$ using Dropout with the VGG (green) and AlexNet (blue) architectures. The mode for Dropout with VGG is $0.04478$ while for the one with AlexNet is $0.04349$. The true value is $\omega_b=0.04492$. }  \label{histofig1}
\end{figure}
\begin{table}[h!]
  \scalebox{0.9}{
\begin{tabular}{|l||l|l|l|l|l|l|l|}
\hline
\multicolumn{7}{|c|}{coverage probabilities from a binned samples }                                                                                                                                                                                                                                          \\ \hline
\multicolumn{1}{|l||}{\backslashbox{C.I}{Model}} & \multicolumn{3}{l|}{\textbf{VGG-Dropout}}& \multicolumn{3}{l|}{\textbf{Alex-Dropout}}  \\ \cline{2-7} 
\multicolumn{1}{|l||}{}                                                                               & $\omega_b$                 & $A_s$  & $\omega_{cdm}$                 & $\omega_b$                 & $A_s$  & $\omega_{cdm}$                     \\ \hline
$68.3\%$                                                                                             & 68.1               &67.1      & 63.3               &68.5             &67.1               & 66.7                         \\ \hline
$95.5\%$                                                                                            & 95.1        &94.5     &93.6           &95.7                   &95.7                   &95.2                               \\ \hline
$99.7\%$                                                                                           &99.6          &99.5     &98.9           &99.8                   &99.6                   &99.4                                    \\ \hline
\end{tabular}}
\caption{\it Estimation of coverage probabilities corresponding to confidence intervals of  $1\sigma$, $2\sigma$ and $3\sigma$. AlexNet was trained with $7\%$ dropout rate, while for VGG we used the network calibrated after training. } \label{table:2appen}
\end{table}9
However,  this distribution  sometimes is  not restricted to be Gaussian, especially for Dropout. In this case, we can follow the method used in~\cite{PerreaultLevasseur:2017ltk} and generate a histogram  from binned samples drawn from the posterior. Since this histogram is expected to be unimodal, we can compute the interval that contains the $(100\alpha)\%$ of the  samples  around the mode, with $\alpha \in[0,1]$.  Fig.~\ref{histofig1} shows the histogram for $\omega_b$ where we can observe the difference by using both architectures, while Table~\ref{table:2appen} reports the coverage probability for individual parameters.   We can observe that the values are  consistent with the ones expected for  a calibrated network.
\section{BNNs Hyper-parameters on the training process}
\label{appendixBhyper}
\begin{figure}[h!]
\begin{center}
\includegraphics[width=0.5\textwidth]{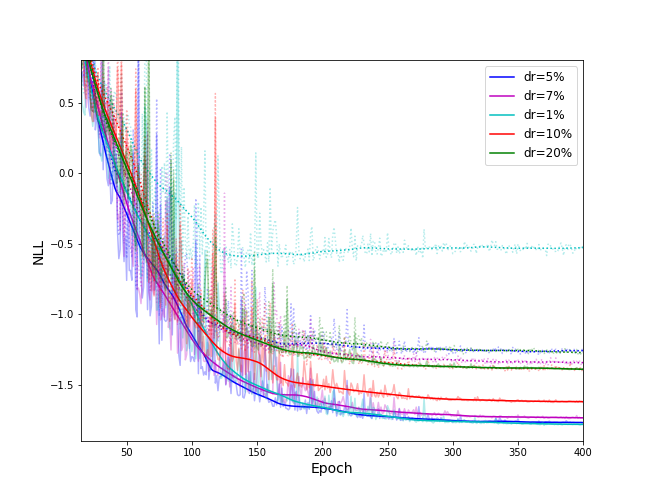}
\end{center}
\caption{\it Training (solid lines) and validation (dotted lines) for AlexNet. Negative log-likelihood for Dropout as a function of the epoch. The colors stand for the used Dropout rate (dr).} \label{fig:4}
\end{figure}

\begin{figure}[h!]
\begin{center}
\includegraphics[width=0.5\textwidth]{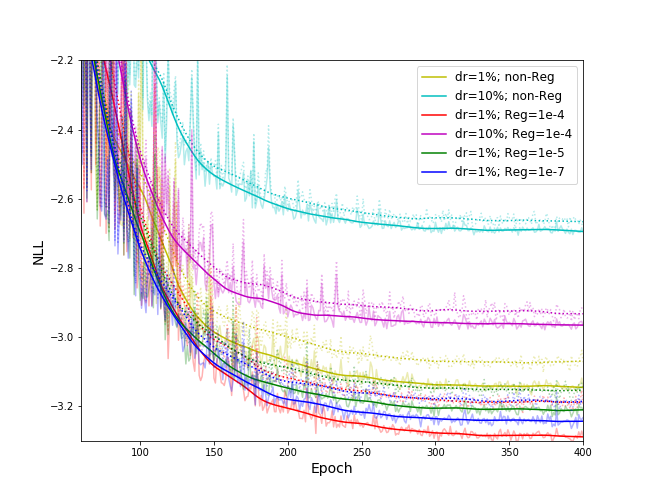}
\end{center}
\caption{\it  Training (solid lines) and validation (dotted lines) Negative log-likelihood for dropout method as a function of epochs. The colors stand for the used dropout rate(dr). The VGG architecture was used for this case.} \label{fig:5}
\end{figure}

\begin{figure}[h!]
\begin{center}
\includegraphics[width=0.5\textwidth]{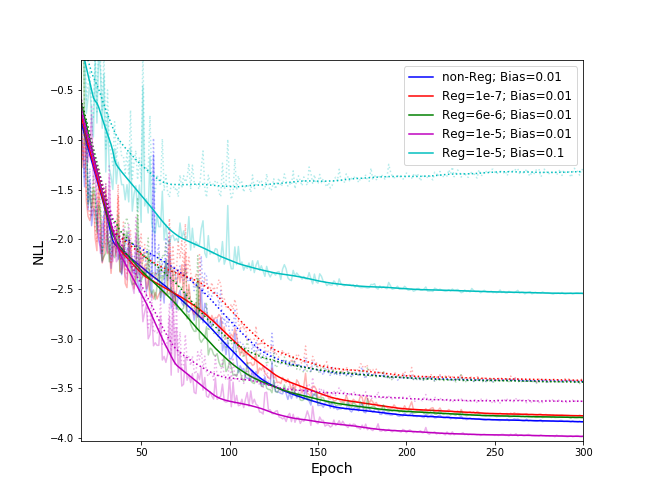}
\end{center}
\caption{\it Training (solid lines) and validation (dotted lines) for AlexNet.
Negative log-likelihood for Flipout method as a function of the epoch. The colors correspond to different regularizers and bias used.} \label{fig:1a}
\end{figure}

Now, to understand the impact of the Dropout rate on the training process, we plot the NLL for different values of the parameter in Fig.~\ref{fig:4} for AlexNet, and  Fig.~\ref{fig:5} for VGG, while   Fig.~\ref{fig:1a} shows the results for AlexNet with Flipout being the regularized the hyper-parameter tuned during training. 

\section{Performance of the BNN using different training dataset sizes}
\label{appendixBla}
\begin{figure}[h!]
\begin{center}
\includegraphics[width=0.52\textwidth]{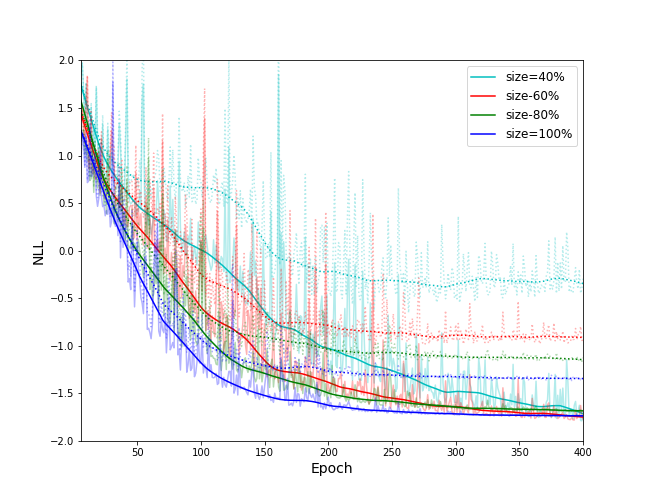}
\end{center}
\caption{\it Training (solid lines) and validation (dotted lines) for AlexNet.
Negative log-likelihood for Dropout method as a function of the epoch. The colors represent the training dataset size  used for a dropout rate of $dr=0.07$. } \label{fig:6}
\end{figure}

In this Appendix we show how the NLL is modified with respect to the training dataset sizes. This kind of analysis not only presents the minimal training dataset size used for obtaining good performance, but also to claim  that epistemic uncertainty can be reduced with sufficient training data~\cite{Gal2015Dropout}.  Fig.~\ref{fig:6} and Fig.~\ref{eq:7} describe the Negative log-likelihood behavior for Dropout and Flipout respectively. 

\begin{figure}[h!]
\begin{center}
\includegraphics[width=0.5\textwidth]{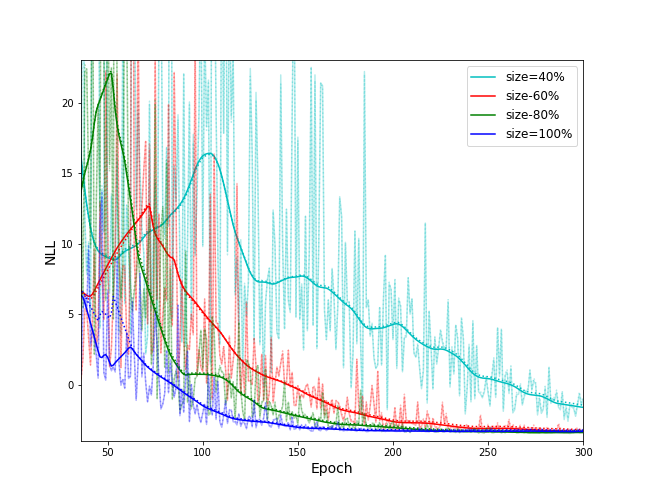}
\end{center}
\caption{\it Training (solid lines) and validation (dotted lines) for VGG. Negative log-likelihood for Flipout method as a function of the epoch. The colors correspond to different training dataset size.} \label{fig:7}
\end{figure}
\section{ Triangle plots for CMB maps from  different BNNs methods}
\label{appendixA}
\begin{figure}[h!]
\begin{center}
\includegraphics[width=0.45\textwidth]{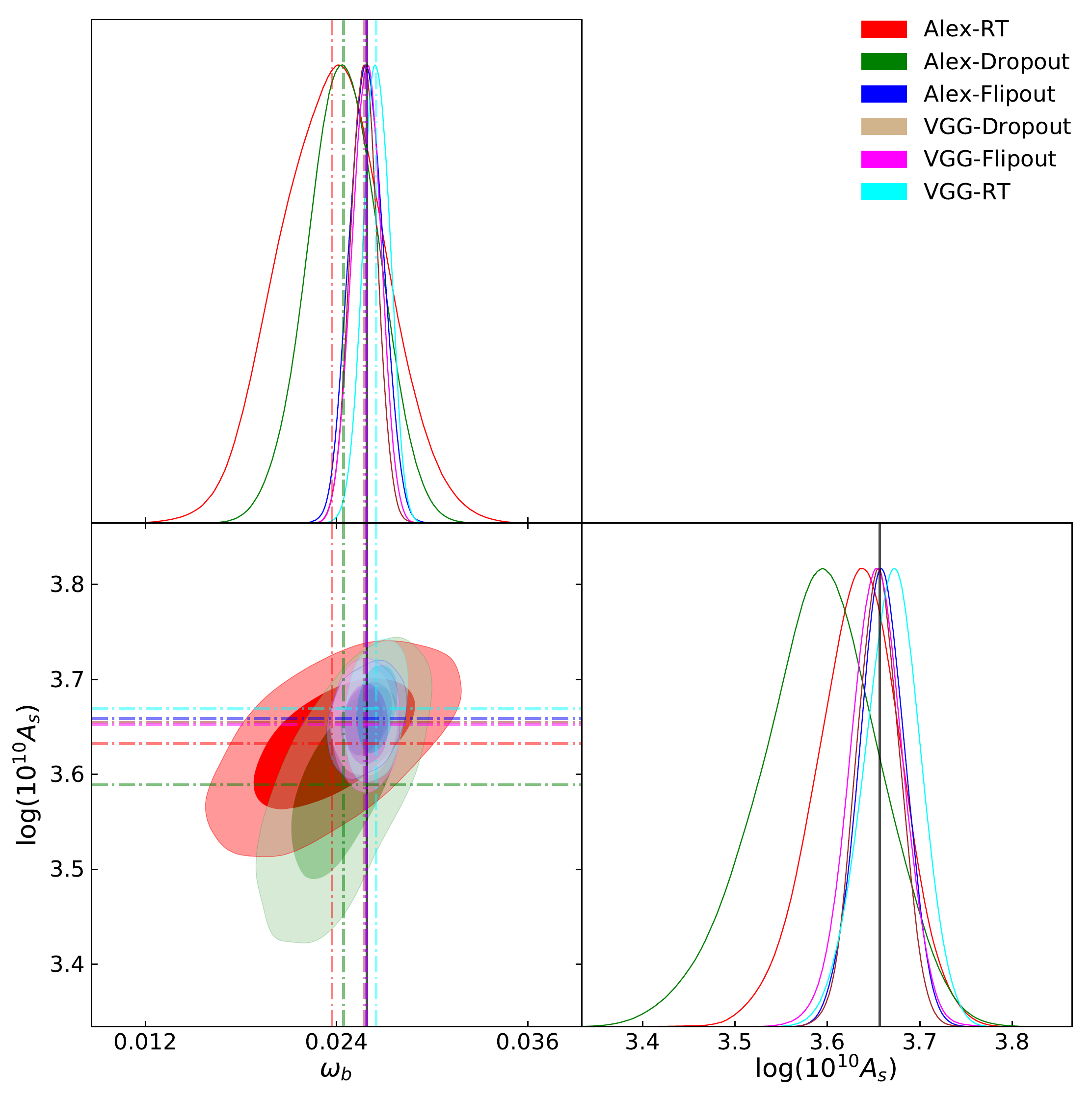}
\end{center}
\caption{\it Minimal base-$\Lambda$CDM $68\%$ and $95\%$ parameter constraint contours from our synthetic CMB dataset using Flipout with Alexnet (blue), VGG (magenta);  Dropout with Alexnet (green), VGG (orange);  and RT with Alexnet (red), VGG (cyan) architectures. The diagonal plots are the marginalized parameter constraints, the dashed lines stand for the predicted values and the black solid line corresponds to the true values $\omega_b=0.02590$, $\log(10^{10}A_s)=3.65653$.}  \label{fig:triang1}
\end{figure}

Fig.~\ref{fig:triang1} displays the results for all BNNs methods introduced in this paper. We can observe that the AlexNet architecture does not work well for RT and Dropout, while for RT we obtained  low performance with respect to Flipout and Dropout. In Table~\ref{table:parameters2} we report the marginalized parameter constraints from the CMB maps. What we can conclude from these results is that the performance for both RT and Dropout depends strongly of the architecture used, as was reported in~\cite{Gal2015BayesianCN}, while for Flipout we do not find this issue. Therefore, Flipout is a more flexible and robust method for obtaining uncertainties at least for CMB dataset.
\begin{table}[h!]
  \scalebox{0.8}{
\begin{tabular}{|l||l|l|l|}
\hline
\multicolumn{3}{|c|}{Marginalized parameter constraints}                                \\ \hline
\multicolumn{1}{|l||}{\backslashbox{$\Lambda$CDM}{BNN}} & \multicolumn{1}{l|}{\textbf{Drop-Alex}}& \multicolumn{1}{l|}{\textbf{RT-Alex}}  \\ \cline{2-3} 
\hline\rule{0pt}{10pt}
$ \omega_{b}$                                           & $0.0245^{+0.0040}_{-0.0039}$ &   $0.0237^{+0.0065}_{-0.0064}$                     \\ \hline
\rule{0pt}{10pt}$\ln(10^{10}A_s)$                                                   & $3.59^{+0.12}_{-0.14}      $ &      $3.633^{+0.088}_{-0.093}   $                                      \\ \hline \rule{0pt}{10pt}
$\omega_{cdm}$                                           & $0.137^{+0.038}_{-0.038}$   &            $0.146^{+0.036}_{-0.037}$ \\ \hline
\end{tabular}}
\caption{\it Parameter $95\%$ intervals for the minimal base-$\Lambda$CDM model from our synthetic CMB dataset using  RT and Dropout with Alexnet. } \label{table:parameters2}
\end{table}
\end{document}